# On new theta identities of fermion correlation functions on genus g Riemann surfaces

A.G. Tsuchiya.

## Abstract

Theta identities on genus g Riemann surfaces which decompose simple products of fermion correlation functions with a constraint on their variables are considered. This type of theta identities is, in a sense, "dual" to Fay's formula, by which it is possible to sum over spin structures of certain part of superstring amplitudes in NSR formalism without using Fay's formula nor Riemann's theta formula in much simpler, more transparent way.    Also, such identities will help to cast correlation functions among arbitrary numbers of Kac-Moody currents in a closed form.    As for genus 1, the identities are reported before in [1] [2].    Based on some notes on genus 1 case which were not reported in [1] [2] and relating those to the results of the Dolan Goddard method [3] on describing Kac-Moody currents in a closed form, we propose an idea of generalizing genus 1 identities to the case of genus g surfaces.    This is not a complete derivation of the higher genus formula due to difficulties of investigating singular part of derivatives of genus g Weierstrass Pe functions.    Mathematical issues remained unsolved for genus g >1 are described in the text.

# 1. Introduction, Conclusions and Notations

What we consider in this document is a simple product of fermion correlation functions

$$S_\delta(z_1, z_2) S_\delta(z_2, z_3) \ldots S_\delta(z_N, z_{N+1}) \quad \text{with a constraint } z_{N+1} = z_1. \quad (1.1)$$

Here $S_\delta(z,w) = \dfrac{\theta[\delta](I(z-w))}{\theta[\delta](0) E(z,w)}$ is the Szego kernel, $E(z,w) = \dfrac{\theta[\nu](I(z-w))}{h_\nu(z) h_\nu(w)}$ is the Prime form of genus g Riemann surfaces, and $I(z-w) \equiv (\int_w^z \omega_i) \in C^g$ represents Abel map. $z_1, z_2, \ldots z_N \in C^1$.

The variables $z_1, z_2, \ldots z_N$ are the inserting points of external particles (bosons) on Riemann surfaces. The constraint $z_{N+1} = z_1$ sometimes naturally occurs in calculating contractions of correlation functions using Wick theorem, and this constraint is a key in our investigations below. Basically, in calculating superstring amplitudes for example, each of the fermion field contractions is multiplied with complicated momentum and polarization vectors of the external particles, and we often face to cope with this type of correlation functions, in the disconnected part of parity-even amplitudes. In summing over spin structures in NSR formalism by applying Fay's formula, one usually faces hard, painful, and untransparent calculations because the formula has a determinant form of fermion correlation functions. The formulae are suitable to calculate free fermion contractions but not in some other cases.

Our aim is to give another type of identities which decompose simple products of Szego kernels into the following form:

$$\prod_i S_\delta(z_i, z_{i+1}) = \sum (\textit{modular invariant functions of } z_i) \cdot (\textit{moduli constant terms}) \quad (1.1)$$

so that the right hand side explicitly shows that spin structure dependence of the product is included only in the moduli constant terms, and to see that the form leads to simpler and more transparent calculations. In general case the key factor in the moduli constant terms is genus g Weierstrass Pe function at non-singular and even half periods of Riemann surfaces, as we will see later. Also, another purpose in this document is to try to obtain explicit form of modular invariant functions of $z_i$ in genus g, which will be remained unaffected in the process of spin sum and which will be contained in the final form of string amplitudes. The method adopted is based on regarding Dolan-Goddard generating function method given in ref. [3] as an expansion by the Weierstrass Pe function. This method leads to a simplified decomposition formula of genus 1 case, as will be shown in **eqs. (2.43)(2.44)(2.45),** and also **(2.47).** We think that the formula of genus g >1 will also be given along the same idea which leads to **eqs. (3.13) (3.14) (3.17),** but there are mathematical difficulties remained unsolved, mainly on structures of genus g sigma function and Pe functions.   If these are fixed, the results may also help describing Kac-Moody currents in a closed form on genus g Riemann surfaces.   The author hopes that someone will accomplish rigorous derivations for general genus along the ideas described here in future.

In this document, we often use matrices and determinants in which the same function is lined up in each column, and the same variable of the functions are lined up in each row:

$$\begin{vmatrix} f_1(x_1) & f_2(x_1) & \cdots & f_N(x_1) \\ f_1(x_2) & f_2(x_2) & \cdots & f_N(x_2) \\ \vdots & \vdots & & \vdots \\ \vdots & \vdots & & \vdots \\ f_1(x_N) & f_2(x_N) & \cdots & f_N(x_N) \end{vmatrix} \tag{1.2}$$

This determinant will be denoted by

$$\det{}_{NxN}[f_1, f_2, \ldots f_N](x_1, x_2, \ldots x_N). \tag{1.3}$$

We also use the following notation, where all elements of the M-th column from the left are replaced with 1,

$$\det{}_{NxN}[f_1, f_2, \ldots f_N](M\,th \to 1; x_1, x_2, \ldots x_N)$$

$$= \begin{vmatrix} f_1(x_1) & \cdots & f_{M-1}(x_1) & 1 & f_{M+1}(x_1) & \cdots & f_N(x_1) \\ f_1(x_2) & \cdots & f_{M-1}(x_2) & 1 & f_{M+1}(x_2) & \cdots & f_N(x_2) \\ \vdots & & & & & & \vdots \\ \vdots & & & & & & \vdots \\ f_1(x_N) & \cdots & f_{M-1}(x_N) & 1 & f_{M+1}(x_N) & \cdots & f_N(x_N) \end{vmatrix} \tag{1.4}$$

An N-th derivative of a function is always denoted as $f^{(N)}(x)$ .

**Notations on genus 1**

Periods:   $\Omega_{m,n} = 2m\omega_1 + 2n\omega_3$

We set $2\omega_1 = 1$, $2\omega_3 = \tau$ in the following.

Half periods $\frac{1}{2}, \frac{\tau}{2}, -\frac{1+\tau}{2}$, denoted by $\omega_\delta$ and $\omega_2 \equiv -(\omega_1 + \omega_3)$, will play important roles.   From sigma-function on torus, we define Pe function as follows:

$$\varsigma(z) = \frac{d\ln\sigma(z)}{dz} \qquad P(z) = -\frac{d\varsigma(z)}{dz}$$

The values of Pe function at half periods are the branch points of the curve:

$$e_\delta = P(\omega_\delta) \qquad (\delta = 1,2,3) \tag{1.5}$$

The branch points $e_\delta$ are related to theta constants as

$$e_1 = \frac{\pi^2(\theta_4^4(0) + \theta_3^4(0))}{3} \quad e_2 = \frac{\pi^2(\theta_2^4(0) - \theta_4^4(0))}{3} \quad e_3 = \frac{-\pi^2(\theta_3^4(0) + \theta_2^4(0))}{3} \tag{1.6}$$

and

$$e_1 + e_2 + e_3 = 0 \qquad e_1 e_2 + e_2 e_3 + e_3 e_1 = -\frac{g_2}{4} \qquad e_1 e_2 e_3 = \frac{g_3}{4} \tag{1.7}$$

Also we use $\quad \eta_\delta = \varsigma(\omega_\delta) \quad (\delta = 1,2,3)$.

The $g_2, g_3$ are classical notations of modular forms which are related to Eisenstein series $G_{2k}(\tau)$ as

$$g_2 = 60 \sum_{m,n} \Omega_{m,n}^{-4} \qquad g_3 = 140 \sum_{m,n} \Omega_{m,n}^{-6} \qquad G_{2k}(\tau) \equiv \sum_{(m,n)\ne(0,0)} \Omega_{m,n}^{-2k} \tag{1.8}$$

We make use of the following relations in the below as well as (1.5):

$$P^{(2n)}(z) = \frac{d^{2n}P(z)}{dz^{2n}} = \text{polynomial of } P(z) \text{ of degree } n+1 \tag{1.9}$$

$$P^{(2n+1)}(z) = P^{(1)}(z) * [\text{polynomial of } P(z) \text{ of degree } n] \tag{1.10}$$

$$P^{(1)}(\omega_\delta) = 0 = P^{(ODD)}(\omega_\delta) \tag{1.11}$$

In particular, the polynomial of (1.9) is denoted as $Q_n(P)$ in the following way.

$$Q_0(P) = 1 \qquad Q_1(P) = P \qquad P^{(2n-2)} = Q_n(P) \tag{1.12}$$

A few examples of $Q_{n+1}(P)$ are:

$$Q_2(P) = P^{(2)} = 6P^2 - \frac{1}{2}g_2 \tag{1.13}$$

$$Q_3(P) = P^{(4)} = 120P^3 - 18g_2 P - 12g_3 \tag{1.14}$$

The $Q_{n+1}(P)$ is consecutively constructed by differentiating

$$\{P^{(1)}(z)\}^2 = 4\{P(z)\}^3 - g_2 P(z) - g_3$$

The coefficient of the highest degree term in $Q_n(P)$ is $(2n-1)!$. $\tag{1.15}$

Two formulae on the fermion correlation function:

$$S_\delta(x) = (P(x) - e_\delta)^{\frac{1}{2}} \quad \text{and} \quad S_\delta(x) = \exp(-\eta_\delta x)\frac{\sigma(x+\omega_\delta)}{\sigma(x)\sigma(\omega_\delta)} \tag{1.16}$$

## 2. Genus 1

On the torus, we discuss

$$\prod_{i=1}^{N} S_\delta(z_i - z_{i+1}), \quad \text{where} \quad S_\delta(z-w) = \frac{\theta_{\delta+1}(z-w)\theta_1^{(1)}(0)}{\theta_{\delta+1}(0)\theta_1(z-w)} \quad \text{and} \quad z_{N+1} = z_1 \quad .$$

$\delta = 1, 2, 3$ corresponds to $R$, $NS$, $\overline{NS}$ respectively.

We define $x_1 \equiv z_1 - z_2$, $x_2 \equiv z_3 - z_2$, ...., $x_N \equiv z_N - z_1$, and hence

$$\sum_{i=1}^{N} x_i = 0. \tag{2.1}$$

As was shown in [2] (originally in [1]), starting from the Frobenius – Stickelberger formula ( $P$ is Weierstrass' Pe function, $P^{(k)}$ are its derivatives. In this document, $F^{(n)}$ always means the differentiation of the function $F$.):

$$\det_{N \times N}[1, P, P^{(1)}, P^{(2)}, P^{(3)}, \ldots P^{(N-2)}](x_1, x_2, \ldots x_N)$$

$$= (-1)^{\frac{(N-1)(N-2)}{2}} 1! 2! \cdots (N-1)! \frac{\sigma(x_1 + x_2 + \cdots x_N) \prod_{\lambda < \mu} \sigma(x_\lambda - x_\mu)}{\prod_{k=1}^{N} \sigma^N(x_k)} \tag{2.2}$$

it is possible to prove the following decomposition formula of product of Szego kernels under the condition $\sum_{i=1}^{N} x_i = 0$ :

$$\prod_{i=1}^{N} S_\delta(x_i) = \frac{1}{(N-1)!} \sum_{K=0}^{\left[\frac{N}{2}\right]} H_{N, N-2K}(x_1, x_2, \ldots x_N) \cdot Q_K(e_\delta) \quad (\text{for each of } \delta = 1, 2, 3) \tag{2.3}$$

where $Q_K(e_\delta)$ is the polynomial defined in (1.12) which appears when re-writing $P^{(2K-2)}$ by $P$ itself and hence $Q_K(e_\delta) = P^{(2K-2)}(\omega_\delta)$ ($P^{(-2)} \equiv 1, P^{(0)} \equiv P, e_\delta = P(\omega_\delta)$).

The constants $e_\delta$ are the branch points of the genus 1 curve which relates to theta constants as in (1.6), and $\omega_\delta$ are the half periods which relates to $e_\delta$ as in (1.5).

The $H_{N,M}$ are manifestly modular invariant functions of $x_i$ written only by Pe function *defined* as follows:

$$H_{N,N} \equiv (-1) \frac{\det_{(N-1) \times (N-1)}[P, P^{(1)}, P^{(2)}, P^{(3)}, \ldots P^{(N-2)}](x_1, x_2, \ldots x_{N-1})}{\det_{(N-1) \times (N-1)}[P, P^{(1)}, P^{(2)}, P^{(3)}, \ldots P^{(N-2)}]((N-1)th \to 1; x_1, x_2, \ldots x_{N-1})}$$

$$\tag{2.4}$$

$$H_{N,N-2K} \equiv + \frac{\det_{(N-1)x(N-1)}[P,\ P^{(1)},P^{(2)},P^{(3)},...... \ P^{(N-2)}]((2K-1)th \to 1; x_1, x_2,...x_{N-1})}{\det_{(N-1)x(N-1)}[P,\ P^{(1)},P^{(2)},P^{(3)},...... \ P^{(N-2)}]((N-1)th \to 1; x_1, x_2,...x_{N-1})}$$

$$(K > 0) \tag{2.5}$$

$$H_{N,0} \equiv 1 \tag{2.6}$$

The summation in (2.3) is until the integer which does not exceed $\frac{N}{2}$. Each of $H_{N,M}$ has the form of a ratio of two determinants, coming from a fact that this is a ratio of two roots of linear equations obtained by Cramer's formula. Note the difference of the overall sign in (2.4) and (2.5). The derivation of eq.(2.3) is described in Appendix A to make this document self-contained.

In the determinants, Pe function and its derivatives are lined up as
$P,\ P^{(1)},P^{(2)},P^{(3)},...... \ P^{(N-2)}$ from the left.
In the denominators, the last, right end column (the $N-1$ th column) is replaced with 1.

For later convenience, we describe one more decomposition formula.
We start from another form of Frobenius – Stickelberger formula:

$$\det_{N \times N}[1,\ P,\ P^{(1)},P^2,\ P^{(1)}P,\ P^3,\ P^{(1)}P^2,...... \ P^{\frac{N}{2}}](x_1, x_2,...x_N)$$

$$= 2^{\frac{N-2}{2}} \frac{\sigma(x_1 + x_2 + \cdots x_N) \prod_{\lambda < \mu} \sigma(x_\lambda - x_\mu)}{\prod_{k=1}^{N} \sigma^N(x_k)}. \tag{2.7}$$

Here, in the determinant, compared with eq.(2.2), even number of derivatives of Pe function $P^{(N)}$ is replaced with monomials $P^{\frac{N}{2}}$, and odd number of derivatives of Pe function $P^{(N)}$ is replaced with monomials $P^{(1)} \cdot P^{\frac{N-1}{2}}$. The order of the poles is $0,2,3,4...N$ from the left, the same as in eq.(2.2).

Then, by the similar argument to derive (2.3) above, we have "another" decomposition formula:

$$\prod_{i=1}^{N} S_\delta(x_i) = \sum_{K=0}^{\left[\frac{N}{2}\right]} V_{N,N-2K}(x_1, x_2,...x_N) \cdot (e_\delta)^K \quad (\textit{for each of } \delta = 1, 2, 3) \tag{2.8}$$

where $V_{N,N-2K}(x_1, x_2,...x_N)$ are obtained by replacing the contents of determinants in

(2.4) (2.5) $P, P^{(1)}, P^{(2)}, P^{(3)}, \ldots P^{(N-2)}$ with

$P, P^{(1)}, P^2, P^{(1)}P, P^3, P^{(1)}P^2, P^{(4)} \ldots$     until $P^{\frac{N}{2}}$ if N is even, $P^{(1)}P^{\frac{N-3}{2}}$ if N is odd. The highest order of poles in this series is N for both cases.

Also, in (2.8),

$$_N V_0 \equiv 1. \tag{2.9}$$

The eq. (2.3) and eq.(2.8) can be written by using half periods as

$$\prod_{i=1}^{N} S_\delta(x_i) = \frac{1}{(N-1)!} \sum_{K=0}^{\left[\frac{N}{2}\right]} H_{N,N-2K}(x_1, x_2, \ldots x_N) \cdot P^{(2K-2)}(\omega_\delta) \tag{2.10}$$

where $P^{(-2)} \equiv 1, P^{(0)} \equiv P$ , and

$$\prod_{i=1}^{N} S_\delta(x_i) = \sum_{K=0}^{\left[\frac{N}{2}\right]} V_{N,N-2K}(x_1, x_2, \ldots x_N) \cdot (P(\omega_\delta))^K \tag{2.11}$$

**Notes:**

1) For the case $N=2$, both of (2.3) and (2.8) becomes

$$S_\delta(x_1) S_\delta(x_2) = -S_\delta^2(x_1) = H_{2,2} + H_{2,0} e_\delta = -P(x_1) + e_\delta, \tag{2.12}$$

with $x_1 + x_2 = 0$.

That is, eq.(2.3) and (2.8), or (2.10) and (2.11), are direct generalizations of a classical formula,

$$S_\delta^2(x) = P(x) - e_\delta .$$

2) For $N=3$, (2.3) and (2.8) is shown to be equivalent to Fay's formula of three variables case. For $N > 3$ it is more convenient to calculate the spin sum than using Fay's formula at genus 1, as explained in [1],[2]. Since the superstring amplitude measures can be written as $\frac{(e_1-e_3)}{\sqrt{D}}, \frac{(e_3-e_2)}{\sqrt{D}}, \frac{(e_2-e_1)}{\sqrt{D}}$ where $\sqrt{D} \equiv \prod_{i<j}^{g}(e_i - e_j)$ , the general form of fermion correlation part becomes, when we use eq.(2.8),

$$\sum_{\delta=1,2,3}(-1)^{\delta}\left[\frac{\theta_{\delta+1}(0)}{\theta_1^{(1)}(0)}\right]^4 \prod_{i=1}^{N} S_{\delta}(x_i) = \sum_{K=0}^{\left[\frac{N}{2}\right]} V_{N,N-2K} \frac{(e_1-e_3)e_2^K + (e_3-e_2)e_1^K + (e_2-e_1)e_3^K}{(e_1-e_3)(e_3-e_2)(e_2-e_1)}$$

(2.13)

The modular invariant functions $V_{N,N-2K}$ remain unaffected under the spin sum, and the moduli dependent terms can be written by elementary symmetric polynomials of $e_1, e_2, e_3$ and hence by modular forms by (1.7).

3) In $V_{N,M}$, $H_{N,M}$, only the terms $N-M \equiv 0 \pmod{2}$ appears as seen in the process of the proof, reflecting the fact that $P^{(1)}(\omega_{\delta}) = 0 = P^{(ODD)}(\omega_{\delta})$.

The $V$, $H$ are given by a ratio of determinants of $(N-1)\times(N-1)$-matrix, with $N-1$ number of variables out of $N$ variables. We can choose any set of $N-1$ variables out of $N$ variables of $x_i$; any choice gives the same $V$, $H$.[1]

4) **Poles and residues**

Define the inverse of our variables $x_i$ of (2.1) as $y_i$ : $y_i \equiv \dfrac{1}{x_i}$.

If we extract out the most singular part of Pe functions from the right hand side of eq.(2.4),(2.5), we have a ratio $\dfrac{L(1,2k,y_1,y_2,\cdots y_{N-1})}{G(1,y_1,y_2,\cdots y_{N-1})}$ for $k=1,2,....$ , where

$$G(1,y_1,y_2,\cdots y_n) \equiv \begin{vmatrix} 1 & y_1^2 & y_1^3 & \cdots & \cdots & y_1^n \\ 1 & y_2^2 & y_2^3 & \cdots & \cdots & y_2^n \\ \vdots & \vdots & \vdots & & & \vdots \\ \vdots & \vdots & \vdots & & & \vdots \\ 1 & y_n^2 & y_n^3 & \cdots & \cdots & y_n^n \end{vmatrix}$$

(B.1)

---

[1] There are two ways to see this fact: First is that the form of $V$, $H$ comes from the ratio of two roots of $N$ number of linear equations of $N-1$ variables, as explained below eq.(A.2). The second is that starting from any $N-1$ numbers in investigating pole structures of $H$, the rest one variable naturally appears and gives the same result, as described at the end of Appendix B.

$$L(1,k,y_1,y_2,\cdots y_n) \equiv \begin{vmatrix} 1 & y_1^2 & y_1^3 & \cdots & y_1^{k-1} & y_1^{k+1} & \cdots & y_1^{n+1} \\ 1 & y_2^2 & y_2^3 & \cdots & y_2^{k-1} & y_2^{k+1} & \cdots & y_2^{n+1} \\ \vdots & \vdots & \vdots & & \vdots & \vdots & & \vdots \\ \vdots & \vdots & \vdots & & \vdots & \vdots & & \vdots \\ 1 & y_n^2 & y_n^3 & \cdots & y_n^{k-1} & y_n^{k+1} & \cdots & y_n^{n+1} \end{vmatrix}$$
(B.2)

In Appendix B, it is proved that, when we denote ${}_nW_m(y_1,y_2,\cdots y_n)$ as the elementary symmetric polynomial of degree m among n variables,

$$\frac{L(1,2k,y_1,y_2,\cdots y_{N-1})}{G(1,y_1,y_2,\cdots y_{N-1})} = {}_NW_{N-2k}(y_1,y_2,\cdots y_N) \quad (2.14)$$

The left hand side includes $N-1$ variables, but on the right hand side includes $N$ variables, due to the condition $\sum_{i=1}^{N} x_i = 0$. Going back to the notations of $x_i$ instead of $y_i$ in ${}_NW_{N-2k}$, we can say that this pole part of Pe function has "simultaneous single pole structure" of $N-2k$ number of variables $b_1, b_2, \ldots b_{N-2k}$ out of $N$ number of variables $x_1, x_2, \ldots x_N$. We denote the elliptic function which has this structure as $[{}_NW_{N-2k}]$.

(Pole structures)

By checking the determinant ratio (2.4)(2.5) carefully, it can be shown that

$$H_{N,M}(x_1,x_2,\ldots x_N) = A_0[{}_NW_M] + A_4[{}_NW_{M-4}] + A_6[{}_NW_{M-6}] + \ldots \quad (2.15)$$

for $N-M = 0 \bmod 2$. In general the constant factors $A_0, A_4, \ldots$ contain the Eisenstein series. See subsection 2-6-2 and the last comment of Appendix D. Note that $H_{N,M}(x_1,x_2,\ldots x_N)$ has not only the first term $A_0[{}_NW_M]$, but also other terms.[2]

The second, third,… terms appear potentially for $N \geq 5$.

(Residues)

In the Appendix B, it is also shown, by considering the ratio of two determinants in (2.14), that if we regard $x_1, x_2, \ldots x_N$ as if they are independent, the residue of $H_{N,M}(x_1,x_2,\ldots x_N)$ satisfies the following equation:

$$\mathop{\mathrm{Res}}_{x_N=0} H_{N,N-2K}(x_1,x_2,\ldots x_N) = H_{N-1,N-2K-1}(x_1,x_2,\ldots x_{N-1}) \quad (2.16)$$

---

[2] On this point ref.[2] has misleading descriptions on the pole structures of the results, although all theta function identities in ref.[2] are correct.

This is already shown before in ref.[3].

**(Re-writing $V_{N,M}$, $H_{N,M}$ in terms of derivatives of sigma function)**

The eq. (2.15) means that the pole of $H_{N,M}$ is of order 1 as of variables $x_1, x_2, \ldots x_N$ under the constraint $\sum_{i=1}^{N} x_i = 0$. This feature matches with the fact that the poles of the product $\prod_{i=1}^{N} S_\delta(x_i)$ is $\dfrac{1}{x_1 x_2 \ldots x_N}$. Utilizing this "single pole" structure, it is possible to re-write the formula of $H_{N,M}, V_{N,M}$ in terms of "derivatives of sigma-function" as follows:

$$N > 1, \quad H_{2,2} = V_{2,2} = P(x_i) \tag{2.17}$$

$$H_{N,0} = V_{N,0} = 1 \tag{2.18}$$

$$H_{N,1} = V_{N,1} = \sum_{i=1}^{N} \zeta(x_i) \tag{2.19}$$

$$c_{21} H_{N,2} = c_{22} V_{N,2} = \left[\sum_{i=1}^{N} \zeta(x_i)\right]^2 - \left[\sum_{i=1}^{N} P(x_i)\right] \tag{2.20}$$

$$c_{31} H_{N,3} = c_{32} V_{N,3} = \left[\sum_{i=1}^{N} \zeta(x_i)\right]^3 - 3\left[\sum_{i=1}^{N} \zeta(x_i)\right]\left[\sum_{i=1}^{N} P(x_i)\right] - \left[\sum_{i=1}^{N} \frac{\partial}{\partial x_i} P(x_i)\right] \tag{2.21}$$

Here $c_{21}, c_{22}, c_{31}, c_{32}$, are numerical constants which will be explicitly determined in subsection 2-6-2, page 18. The minus signs on the right hand sides of (2.20),(2.21) are from the definition of Pe function. All of the $H_{N,M}$ in eqs.(2.19)(2.20)(2.21) have the pole structures of (2.15) for M=1,2, 3. In investigating superstring amplitudes, this form is sometimes more convenient than the determinant-ratio formula (2.4),(2.5), (2.8). It is reported in ref.[3] that naive extensions of the results of (2.19)-(2.21) to higher values of M for $H_{N,M}$ are not working well. This fact closely relates to the existence of the second, third, … terms of the right hand side of eq.(2.15) on the pole structures of $H_{N,M}$ in general. However there is a remedy on this, and we can derive a formula for any $H_{N,M}$. See subsection 2-6-2 of this document.

Alternatively, in ref.[4], a way of constructing elliptic functions as derivatives of sigma function which have simultaneous single poles in the variables $x_1, x_2, \ldots x_N$ for arbitrary $N$ is reported, in the context of elliptic multiple zeta values.  Since such functions are essentially determined by the pole structures, those functions would be the same as $H_{N,M}$ or $V_{N,M}$ up to moduli dependent constant terms.  It is certain that the product of Szego kernels $\prod_{i=1}^{N} S_\delta(x_i)$ can be expanded by $V^{eMZV}{}_{N-2K}(x_1, x_2, \ldots x_N)$.  Moreover, the method is applied to non-maximal super symmetric case in ref.[5].  On discussions related to the structures of superstring amplitude structures, see also [6][7].  [**Note added:** The generating function method which will be described in section 2-6-2 below and the elliptic multiple zeta values method for the genus one spin sum were found out to be equivalent.  Appendix E is added for explaining it in the last version of this document. ]

### 5) Zeros

The even theta functions at genus 1, $\theta_{\delta+1}(x)$, have zero points at $\omega_\delta$. That is,

$\theta_2(x)$, $\theta_3(x)$, $\theta_4(x)$ becomes zero if $x$ equals to half periods

$$\frac{1}{2} = \omega_1, \frac{1+\tau}{2} = \omega_2, \frac{\tau}{2} = \omega_3, \qquad (2.22)$$

respectively.  Since the fermion correlation function has the form

$S_\delta(z-w) = \dfrac{\theta_{\delta+1}(z-w)\theta_1^{(1)}(0)}{\theta_{\delta+1}(0)\theta_1(z-w)}$, the product $\prod_{i=1}^{N} S_\delta(x_i)$ becomes zero if *any* of the variables $x_i$ is equal to $\omega_\delta$. Therefore, the right hand side of (2.3), (2.8) should also has this feature.    This can be checked in general in the process of proving (2.3). Here we exemplify one example of N=4.

$$\prod_{i=1}^{4} S_\delta(x_i) = -\frac{\begin{vmatrix} P(x_1) & P^{(1)}(x_1) & P^2(x_1) \\ P(x_2) & P^{(1)}(x_2) & P^2(x_2) \\ P(x_3) & P^{(1)}(x_3) & P^2(x_3) \end{vmatrix}}{\begin{vmatrix} 1 & P(x_1) & P^{(1)}(x_1) \\ 1 & P(x_2) & P^{(1)}(x_2) \\ 1 & P(x_3) & P^{(1)}(x_3) \end{vmatrix}} + \frac{\begin{vmatrix} 1 & P^{(1)}(x_1) & P^2(x_1) \\ 1 & P^{(1)}(x_2) & P^2(x_2) \\ 1 & P^{(1)}(x_3) & P^2(x_3) \end{vmatrix}}{\begin{vmatrix} 1 & P(x_1) & P^{(1)}(x_1) \\ 1 & P(x_2) & P^{(1)}(x_2) \\ 1 & P(x_3) & P^{(1)}(x_3) \end{vmatrix}} e_\delta + e_\delta^2 ,$$

(2.23)

When $x_1 = \omega_\delta$, since $P(\omega_\delta) = e_\delta$ and $P^{(1)}(\omega_\delta) = 0$,

Numerator =

$$-\begin{vmatrix} e_\delta & 0 & e_\delta^2 \\ P(x_2) & P^{(1)}(x_2) & P^2(x_2) \\ P(x_3) & P^{(1)}(x_3) & P^2(x_3) \end{vmatrix} + \begin{vmatrix} 1 & 0 & e_\delta^2 \\ 1 & P^{(1)}(x_2) & P^2(x_2) \\ 1 & P^{(1)}(x_3) & P^2(x_3) \end{vmatrix} e_\delta + \begin{vmatrix} 1 & e_\delta & 0 \\ 1 & P(x_2) & P^{(1)}(x_2) \\ 1 & P(x_3) & P^{(1)}(x_3) \end{vmatrix} e_\delta^2$$

$$=0 \qquad (2.24)$$

If, one of the variables of variables $x_i$ is set equal to $\omega_\delta$ in (2.17)-(2.20) or in general formula (2.45), it leads to non-trivial theta identities, but those may not be so interesting.

### 6) Generating function method in ref. [3] : its modification and applications

In ref.[3], Dolan-Goddard introduced a generating function $H^{DG}_N$ so that coefficients $H^{DG}_{N,M}$ have the desired residue structures which is the same as in (2.16), to obtain N point one loop correlation functions for the currents of an arbitrary affine Kac-Moody algebra in a closed form:

$$v^{-N} H^{DG}_N = \prod_{i=1}^{N} \frac{\sigma(z_{i+1} - z_i + v)}{\sigma(z_{i+1} - z_i)\sigma(v)} = \prod_{i=1}^{N} \frac{\sigma(x_i + v)}{\sigma(x_i)\sigma(v)} = \sum_{M=0}^{\infty} H^{DG}_{N,M}(x_i) v^{M-N} \qquad (2.25)$$

The $H^{DG}_{N,M}(x_i)$ are defined in this equation as coefficients of the expansion of a function $\prod_{i=1}^{N} \frac{\sigma(x_i + v)}{\sigma(x_i)\sigma(v)}$.

This form of equation reminds us a classical formula on fermion correlation function on torus:

$$S_\delta(x) = \exp(-\eta_\delta x) \frac{\sigma(x + \omega_\delta)}{\sigma(x)\sigma(\omega_\delta)} \qquad (2.26)$$

Then

$$\prod_{i=1}^{N} S_\delta(x_i) = \prod_{i=1}^{N} \frac{\sigma(x_i + \omega_\delta)}{\sigma(x_i)\sigma(\omega_\delta)} \qquad (2.27)$$

because the exp factor cancels by the condition $\sum_{i=1}^{N} x_i = 0$.

To pursue this similarity further, we consider matching the singular terms of (2.25) and Pe function:

$$P(v) = v^{-2} + \sum_{m=1}^{\infty} c_{2m} v^{2m}, \quad P^{(n)}(v) = \frac{d_n}{v^{n+2}} + n!c_n + \ldots\ldots, \quad d_n = (-1)^n (n+1)! \quad (2.28)$$

Here, $c_{2m}$ is related to Eisenstein series as $c_{2m} = (2m+1)G_{2m+2}(\tau)$.

Writing the right hand side of eq.(2.25) as

$$\prod_{i=1}^{N} \frac{\sigma(x_i + v)}{\sigma(x_i)\sigma(v)} = \sum_{M=0}^{\infty} H^{DG}_{N,M}(x_i) v^{M-N} = H^{DG}_{N,0} \frac{1}{v^N} + H^{DG}_{N,1} \frac{1}{v^{N-1}} + \ldots\ldots \quad (2.29)$$

and comparing with the singular part of derivatives of Pe, $\frac{1}{v^N}, \frac{1}{v^{N-1}}, \ldots\ldots$ we have

$$\prod_{i=1}^{N} \frac{\sigma(x_i + v)}{\sigma(x_i)\sigma(v)} = \frac{H^{DG}_{N,0}}{d_{N-2}} P^{(N-2)}(v) + \frac{H^{DG}_{N,1}}{d_{N-3}} P^{(N-3)}(v) + \frac{H^{DG}_{N,2}}{d_{N-4}} P^{(N-4)}(v) + \ldots\ldots (2.30)$$

This matching ends at finite time; if all singular terms and the constant term match, then the holomorphic part will match automatically. Therefore we have an expansion by the Pe function as

$$\prod_{i=1}^{N} \frac{\sigma(x_i + v)}{\sigma(x_i)\sigma(v)} = \sum_{M=0}^{N} \frac{H^{DG}_{N,M}}{d_{N-2-M}} P^{(N-2-M)}(v). \quad (2.31)$$

$(\ P^{(-2)} \equiv 1 \quad \text{and} \quad P^{(0)} \equiv P \ )$

with $H^{DG}_{N,N-1} = 0$ since there is no single pole in Pe, and $\frac{H^{DG}_{N,0}}{d_{N-2}} = 1$. Note that $H^{DG}_{N,N}$ on which no singular function of $v$ is multiplied is *defined* in (2.31) in this document.

Since $P^{(odd)}(\omega_\delta) = 0$, if we set $v = \omega_\delta$, this becomes

$$\prod_{i=1}^{N} \frac{\sigma(x_i + \omega_\delta)}{\sigma(x_i)\sigma(\omega_\delta)} = \prod_{i=1}^{N} S_\delta(x_i) = \sum_{K=0}^{\left[\frac{N}{2}\right]} \frac{H^{DG}_{N,N-2K}}{d_{2K-2}} P^{(2K-2)}(\omega_\delta). \quad (2.32)$$

That is, $H^{DG}_{N,M}$ relates to our $H_{N,M}$ in eq.(2.4) (2.5) which were originally obtained as a closed form as a ratio of two determinants by a trick explained in Appendix A without considering any expansions like (2.30). We have

$$\frac{H_{N,M}(x_1, x_2, \ldots x_N)}{(N-1)!} = \frac{H^{DG}_{N,M}(x_1, x_2, \ldots x_N)}{d_{N-M-2}} \quad (2.33)$$

for the terms $N - M \equiv 0 \pmod 2$. The generating function method is equivalent to the expansion by the derivatives of Pe function as

$$\prod_{i=1}^{N} \frac{\sigma(x_i + v)}{\sigma(x_i)\sigma(v)} = \frac{1}{(N-1)!} \sum_{M=0}^{N} H_{N,M}(x_i) P^{(N-2-M)}(v). \quad (2.34)$$

Two notes, which we will consider again later when we try to generalize the identities to genus g case in chapter 3:

**2-6-1:**

Consider an expansion of a function $L(z_i, \nu)$, whose variables are vertex inserting points $z_i$ and a sub-variable $\nu$, by the derivatives of Pe function with some coefficients $\overline{H}_{N,N-M}(z_i)$ as

$$L(z_i, \nu) = \sum_{M=0}^{N} \overline{H}_{N,N-M}(z_i) P^{(N-2-M)}(\nu) \qquad (P^{(-2)} \equiv 1 \text{ and } P^{(0)} \equiv P) \qquad (2.35)$$

$L(z_i, \nu)$ should have singular part of $\nu$, and $N$ is some number determined by the highest order of the pole $\nu$.

Also consider one more expansion of the function by first derivative of Pe function and monomials of Pe as

$$L(z_i, \nu) = \overline{V}_{N,N} + \overline{V}_{N,N-2}(z_i) P + \overline{V}_{N,N-3}(z_i) P^{(1)} + \overline{V}_{N,N-4}(z_i) P^2 + \overline{V}_{N,N-5}(z_i) P^{(1)} P + \ldots$$

(2.36)

It is always possible to do these expansions (2.35), (2.36) as long as $L(z_i, \nu)$ is elliptic, since the first derivative of Pe function and monomials of Pe, which relate to higher derivatives of Pe function, are the basis of the function space at genus 1.

As can be seen from (2.35) and (2.36), if we consider some objects which have the form $L(z_i, \omega_\delta)$ in super string amplitudes, since $P^{(ODD)}(\omega_\delta) = 0$, the spin structure dependence of $L(z_i, \omega_\delta)$ is only through one kind of constants $P(\omega_\delta)$ which, in genus one, equals to the branch point $e_\delta$. One important condition to be imposed on $L(z_i, \nu)$ is that it should be modular invariant so that the expansion coefficients are also modular invariant.

The contraction $\prod_{i=1}^{N} S_\delta(x_i)$ is one of such cases, by the condition $z_{N+1} = z_1$, where the function $L(z_i, \nu)$ is $\prod_{i=1}^{N} \frac{\sigma(x_i + \nu)}{\sigma(x_i)\sigma(\nu)}$. The $\nu$ dependent function $\sigma(\nu)$ in the denominator is necessary to have singularities and be expanded by Pe function. This is a natural explanation why we can decompose $\prod_{i=1}^{N} S_\delta(x_i)$ into the form (2.3) or (2.8), where spin structure dependence is only through $P(\omega_\delta)$. It would be also natural to extend this observation to the case of general genus.

## 2-6-2 On expressing $H_{N,M}(x_i)$ by polynomials of derivatives of sigma function

The function $\prod_{i=1}^{N} \frac{\sigma(x_i + \nu)}{\sigma(x_i)\sigma(\nu)}$ can be used as a generating function, to obtain the form of $H_{N,M}$, as follows. First, if we differentiate the function U defined as

$$U(\nu) \equiv \exp[\sum_{i=1}^{N} \ln \sigma(x_i + \nu) - \sum_{i=1}^{N} \ln \sigma(x_i)] \quad . \tag{2.37}$$

with respect to $\nu$ up to $M$ times and set $\nu$ equal to zero, then the differentiations with respect to $\nu$ change into differentiation with respect to $x_i$ because the combination of $x_i$ and $\nu$ is linear. We can obtain polynomials of derivatives of sigma function which have "simultaneous single pole structure of $M$ number of variables" out of $N$ variables $x_1, x_2,....x_N$, for any $M$. That is, if we choose $M$ variables out of the variables $x_1, x_2,....x_N$ then the poles of $\frac{\partial^M}{\partial \nu^M} U(\nu)\big|_{\nu=0}$ are all combinations of the inverse of the products of such $M$ variables. This can be proved without difficulty by mathematical induction; see Appendix D. It is straightforward to see that this procedure reproduces all of (2.18), (2.19), (2.20), (2.21) for $M =1,2,3$.

However this naïve procedure does not work for $M \geq 4$ to obtain $H_{N,M}(x_i)$ in general. The U is equal to $\prod_{i=1}^{N} \frac{\sigma(x_i + \nu)}{\sigma(x_i)}$ by the definition (2.37), which is different from $\prod_{i=1}^{N} \frac{\sigma(x_i + \nu)}{\sigma(x_i)\sigma(\nu)}$. The procedure here is already described in ref.[3] chapter 3 in a different context; the U is the same as H_hat of eq.(3.63) in ref.[3]. In that article, the correct form of generating function $\prod_{i=1}^{N} \frac{\sigma(x_i + \nu)}{\sigma(x_i)\sigma(\nu)}$ was found out from the residue point of view about the modular invariant functions $H_{N,M}(x_i)$. We saw that it relates to the product of Szego kernels by the formula (2.26) with the constraint $\sum_{i=1}^{N} x_i = 0$. Obtaining these polynomials for arbitrary $M$ from U in this way is still very useful as we will see below, but not enough to derive general formulae of $H_{N,M}$ nor to derive a general procedure of spin sums.

Instead we adopt to do as follows. Express a function $[\frac{\nu}{\sigma(\nu)}]^N \prod_{i=1}^{N} \frac{\sigma(x_i+\nu)}{\sigma(x_i)}$ in the following form, by using a formula ( infinite product representation of sigma function)

$$\sigma(\nu) = \nu \cdot \prod_{m,n}(1-\frac{\nu}{\Omega_{m,n}}) \exp[\frac{\nu}{\Omega_{m,n}}+\frac{\nu^2}{2\Omega_{m,n}^2}] \tag{2.38}$$

as

$$[\frac{\nu}{\sigma(\nu)}]^N \prod_{i=1}^{N} \frac{\sigma(x_i+\nu)}{\sigma(x_i)} = \exp[\sum_{i=1}^{N}\ln\sigma(x_i+\nu) - \sum_{i=1}^{N}\ln\sigma(x_i) - N\sum_{m,n}[\ln(1-\frac{\nu}{\Omega_{m,n}}) + (\frac{\nu}{\Omega_{m,n}}+\frac{\nu^2}{2\Omega_{m,n}^2})]] \tag{2.39}$$

If we differentiate $N\sum_{m,n}[\ln(1-\frac{\nu}{\Omega_{m,n}}) + (\frac{\nu}{\Omega_{m,n}}+\frac{\nu^2}{2\Omega_{m,n}^2})]$ one, two, three times it becomes zero after setting $\nu \to 0$. If differentiate four times or more, it always gives one monomial of Eisenstein series $\sum_{m,n} \frac{1}{\Omega_{m,n}^k}$ as $\nu \to 0$. This is nothing but the same as Laurent expansion process of Pe-function as a series of $\nu$ excluding its singular part. When differentiate 4 times or more, the terms $(\frac{\nu}{\Omega_{m,n}}+\frac{\nu^2}{2\Omega_{m,n}^2})$ are not important because they vanish. If we always differentiate these terms finite times with respect to $\nu$ and then put $\nu \to 0$, by using

$$\frac{\partial^k}{\partial \nu^k}\ln(1-\frac{\nu}{\Omega_{m,n}})\Big|_{\nu=0} = -\frac{(k-1)!}{\Omega_{m,n}^k}, \quad \text{the terms } N\sum_{m,n}[\ln(1-\frac{\nu}{\Omega_{m,n}}) + (\frac{\nu}{\Omega_{m,n}}+\frac{\nu^2}{2\Omega_{m,n}^2})]$$

can be replaced with $-N\sum_{k=2}^{\infty}\frac{1}{2k}G_{2k}(\tau)\nu^{2k}$, where $G_{2k}(\tau) \equiv \sum_{(m,n)}\frac{1}{(m+n\tau)^{2k}}$ is the Eisenstein series.

We write (2.34) by multiplying $\nu^N$ to both sides as

$$[\frac{\nu}{\sigma(\nu)}]^N \prod_{i=1}^{N}\frac{\sigma(x_i+\nu)}{\sigma(x_i)} = \frac{\nu^N}{(N-1)!}\sum_{M=0}^{N}H_{N,M}(x_i)P^{(N-2-M)}(\nu). \tag{2.40}$$

Due to simple pole structure of Pe function:

$$v^N P^{(N-2-M)}(v) = d_{N-2-M} v^M + O(v^N). \quad , \quad d_n = (-1)^n (n+1)!, \tag{2.41}$$

there are no singular terms on both sides of (2.40). To get the form of $H_{N,M}(x_i)$ for arbitrary $N$ and $M$, we only have to differentiate both sides of (2.40) $M$ times and set $v = 0$. For $M < N$, instead of differentiating eq.(2.37), we have

$$H_{N,M}(x_i) = \frac{(-1)^{N-M}(N-1)!}{(N-1-M)!M!} \frac{\partial^M}{\partial v^M} \exp[\sum_{i=1}^{N} \ln\sigma(x_i + v) - \sum_{i=1}^{N} \ln\sigma(x_i) + N\sum_{k=2}^{\infty} \frac{1}{2k} G_{2k}(\tau) v^{2k}]\bigg|_{v=0}$$
$$(for \quad M < N) \tag{2.42}$$

The existence of the last term in the exp function which contains Eisenstein series affects the polynomial form of the derivatives of sigma function. The poles of $H_{N,M}(x_i)$ are not only the combinations of inverse of exactly $M$ number of variables from $N$ variables $x_i$, but also those of less numbers, for $M > 4$. This fact relates to the structure of eq.(2.15), and also reflects what was pointed out in ref.[3] on the residues.

When $M = N$, before differentiating N times, we should pay attention because the right hand side of (2.40) has plural terms proportional to $v^N$. We have, as described in Appendix D,

$$H_{N,M}(x_i) = \frac{(-1)^{N-M}(N-1)!}{(N-1-M)!M!} \frac{\partial^M}{\partial v^M} \exp[\sum_{i=1}^{N} \ln\sigma(x_i + v) - \sum_{i=1}^{N} \ln\sigma(x_i) + N\sum_{k=2}^{\infty} \frac{1}{2k} G_{2k}(\tau) v^{2k}]\bigg|_{v=0}$$
$$- \sum_{K=2}^{N-2} \{(K+1)! G_{K+2}(\tau) H_{N,N-K-2}(x_i)\} \delta_{N,M} \tag{D.4}$$

The Eisenstein series $G_{K+2}(\tau)$ are zero if $K+2$ is odd.
We modify (D.4) as

$$H_{N,M}(x_i) = \frac{\partial^M}{\partial v^M} F(v) \bigg|_{v=0} \tag{2.43}$$

where

$$F(v) \equiv \frac{(-1)^{N-M}(N-1)!}{(N-1-M)!M!} \exp[\sum_{i=1}^{N} \ln\sigma(x_i + v) - \sum_{i=1}^{N} \ln\sigma(x_i)$$
$$+ N\sum_{k=2}^{\infty} \frac{1}{2k} G_{2k}(\tau) v^{2k} - \frac{1}{(N-1)!} v^N \sum_{K=2}^{N-2} \{(K+1)! G_{K+2}(\tau) H_{N,N-K-2}(x_i)] \tag{2.44}$$

Eq. (2.43) is equivalent to the determinant-ratio formula (2.4) (2.5), and represents the general form of modular invariant functions of vertex inserting points. The factor $[\frac{v}{\sigma(v)}]^N$ comes from the denominator of $\prod_{i=1}^{N} \frac{\sigma(x_i + v)}{\sigma(x_i)\sigma(v)}$, and it does not depend on the differences of vertex inserting points or $x_i$, so, if this factor contributes to the modular invariant function $H_{N,M}(x_i)$ the result should be constant modular forms.

On numerical factors: $(N-1)!$ in (2.40) and (2.44) comes from the process of proof of (2.3); this is the coefficients of highest degree of Pe when $P^{(N-2-M)}(v)$ is written as a polynomial of Pe. The factor $M!$ in (2.44) is from $\dfrac{\partial^M}{\partial v^M}$ to derive H. The factor $\dfrac{(-1)^{N-M}}{(N-1-M)!}$ in (2.44) is from $d_n = (-1)^n (n+1)!$ in (2.41).

Then, by (2.3) or (2.10), or by setting $v = \omega_\delta$ in (2.34),

$$\prod_{i=1}^{N} S_\delta(x_i) = \frac{1}{(N-1)!} \sum_{K=0}^{\left[\frac{N}{2}\right]} H_{N,N-2K}(x_1, x_2, \ldots x_N) \cdot P^{(2K-2)}(\omega_\delta)$$

$$= \frac{1}{(N-1)!} \sum_{K=0}^{\left[\frac{N}{2}\right]} \frac{\partial^{N-2K}}{\partial v^{N-2K}} F(v)\Big|_{v=0} P^{(2K-2)}(\omega_\delta) \qquad (2.45)$$

$$(-1)! \equiv (0)! \equiv 1, \quad M \equiv N - 2K$$

As noted, all of the differentiations with respect to $v$ change to differentiations with respect to $x_i$ by setting $v = 0$. This result (2.45), as well as (2.43), includes theta identities we need practically at genus 1 after all.

The derivatives of sigma function are expressed by derivatives of $\ln \theta_1(x_i)$, where $\theta_1(x)$ is unique odd theta function on torus. An extra constant $\eta_1$ appears only when we differentiate $\ln \theta_1(x_i)$ two times, in $\sum_{i=1}^{N} P(x_i)$.

Eq.(2.43) also says that the ratios of two determinants (2.4) (2.5) (2.6) have expansion formula as in the right hand side of (2.43), when $M = N - 2K$.

If we adopt (2.45), the following is the closed form of the contractions of fermion field in one loop N point superstring amplitudes of parity conserving part after the spin structure sum:

$$\sum_{\delta=1,2,3} (-1)^\delta \left[\frac{\theta_{\delta+1}(0)}{\theta_1^{(1)}(0)}\right]^4 \prod_{i=1}^{N} S_\delta(x_i)$$

$$= \frac{1}{(N-1)!} \sum_{K=0}^{\left[\frac{N}{2}\right]} \frac{\partial^{N-2K}}{\partial v^{N-2K}} F(v)\Big|_{v=0} \frac{(e_1-e_3)Q_K(e_2) + (e_3-e_2)Q_K(e_1) + (e_2-e_1)Q_K(e_3)}{(e_1-e_3)(e_3-e_2)(e_2-e_1)}$$

(2.46)

where we re-wrote $P^{(2K-2)}$ by $P$ itself and hence $Q_K(e_\delta) = P^{(2K-2)}(\omega_\delta)$

$(P^{(-2)} \equiv 1, P^{(0)} \equiv P \quad e_\delta = P(\omega_\delta), \quad M \equiv N - 2K)$.

As shown in ref[1], or explicitly shown in ref.[2] for arbitrary $N$, the spin structure sum reduces to elementary algebras of branch points $e_i$ in the $Q_K(e_\delta)$ dependent factor $\dfrac{(e_1 - e_3)Q_K(e_2) + (e_3 - e_2)Q_K(e_1) + (e_2 - e_1)Q_K(e_3)}{(e_1 - e_3)(e_3 - e_2)(e_2 - e_1)}$ of eq.(2.46).

By the concrete forms of polynomials of $Q_K(e_\delta)$ given in (1.12), (1.13), (1.14), terms for K=0,1,3 are zero. In general the factor can be represented by polynomials of elementary symmetric functions of $e_1, e_2, e_3$ and consequently by the Eisenstein series $G_4(\tau)$ and $G_6(\tau)$, via eq.(1.7) and eq.(1.8).

For any hyper elliptic cases, the spin structure sum will be done in the same manner, and spin structure dependent terms will be always represented by the symmetric polynomials of branch points and hence by modular forms after the summations. This higher genus fact is quite plausible but not yet proved in general [2].

Note that the Eisenstein series appear [6][7] from two places. One is in the function on the exponential factor of $F(\nu)$, coming from elliptic functions of vertex inserting points. The other is from the $Q_K(e_\delta)$ dependent factor. The former contains $G_4(\tau)$ at the N=6 level, which appears in Kac-Moody currents closed form as $k_4$ of ref.[3] eq(3.52), but in string amplitudes this term disappears after the spin structure sum because it corresponds to N=6, K=1 term in (2.46).

Also, note that, since the terms of K=0,1,3 do not contribute to the result, the results of the summation of eq.(2.46) do not contain $H_{N,N}(x_i)$, $H_{N,N-2}(x_i)$, $H_{N,N-6}(x_i)$. In particular, since $H_{N,N}(x_i)$ does not contribute, the last terms in (2.44) proportional to $\nu^N$ can be disregarded, or equivalently we can use eq.(2.42) for the expression of $H_{N,M}(x_i)$ in the spin sum. Since the contractions of the boson fields in the vertex operators do not give particle poles in superstring amplitudes, these functions represent the pole structures of the differences of vertex inserting points in one loop amplitudes in general.

For $N \leq 4$, algebras of $e_\delta$ dependent terms in eq.(2.46) give only non-zero result when $N = 4$ and $K = 2$. That non-zero result is 1. This fact includes non-renormalization theorems for $N \leq 3$ and four point amplitude results of superstrings.

For $N \leq 7$, the $Q_K(e_\delta)$ dependent factor always gives 0 or numerical constants for any of K.

At $N = 8$, $Q_K(e_\delta)$ dependent factor gives first non-trivial modular form, as explicitly described in ref.[4]. This is from $N - 2K \geq 0$ and $K \geq 4$, since $K = 3$ gives zero for $Q_K(e_\delta)$ dependent factor.

And also for $N \geq 8$, Eisenstein series in the factor $F(\nu)$ contributes non-zero results in string amplitudes. This is from $N - 2K \geq 4$ and $K \geq 2$.

On the other hand, the functions of $x_i$, differences of vertex inserting points, are always simply come from the differentiations of sigma functions on the exponential factor $F(\nu)$. This fact may be a hint to consider higher genus cases because even in higher genus the pole structures as for the differences of vertex inserting points are the same.

Considering all above, the eq.(2.46) has an expression as

$$\sum_{\delta=1,2,3} (-1)^\delta \left[\frac{\theta_{\delta+1}(0)}{\theta_1^{(1)}(0)}\right]^4 \prod_{i=1}^N S_\delta(x_i) = \left[\frac{1}{(N-4)!}\frac{\partial^{N-4}}{\partial \nu^{N-4}} + \sum_{K=4}^{\left[\frac{N}{2}\right]} \frac{E_K(\tau)}{(N-2K)!}\frac{\partial^{N-2K}}{\partial \nu^{N-2K}}\right] F_N(x_i, \nu)\bigg|_{\nu=0}$$

(2.47)

where

$$E_K(\tau) \equiv \frac{1}{(2K-1)!} \frac{(e_1 - e_3)Q_K(e_2) + (e_3 - e_2)Q_K(e_1) + (e_2 - e_1)Q_K(e_3)}{(e_1 - e_3)(e_3 - e_2)(e_2 - e_1)}$$

$$F_N(x_i, \nu) \equiv \exp\left[\sum_{i=1}^N \ln\sigma(x_i + \nu) - \sum_{i=1}^N \ln\sigma(x_i) + N\sum_{k=2}^\infty \frac{1}{2k} G_{2k}(\tau)\nu^{2k}\right] \quad (2.48)$$

which is the same as $H_{N,M}(x_i)$ in (2.42) except the over-all numerical factor. The last term in (2.47), in which the summation begins at $K = 4$, is non-zero only when $N \geq 8$.

As commented above, in eq.(2.48), we do not have to use the full form of $F(\nu)$ of eq.(2.44) because $H_{N,N}(x_i)$ is not included. There is no ambiguity, especially on adding or multiplying modular invariant constant terms or factors, in the process of deriving (2.47), (2.48) starting from eq.(2.34), for any value of $N$.

The first term of the right hand side of eq.(2.47), $\frac{1}{(N-4)!} \frac{\partial^{N-4}}{\partial v^{N-4}} F_N(x_i, v)\big|_{v=0}$, reproduces all of spin sum results up to $N \leq 7$ in ref.[1]. The $K = 4,5,6$ terms in the second term of eq.(2.47) with the first term should match with the results of ref.[4] up to $N \leq 12$. The spin sum method and results were described in ref.[2], where the modular invariant functions of $x_i$ were expressed by "determinant-ratio formula" of (2.10) or (2.11). In this document, those $x_i$ dependent terms are re-written by the derivatives of sigma function and the Eisenstein series directly as $\frac{\partial^{N-2K}}{\partial v^{N-2K}} F_N(x_i, v)\big|_{v=0}$. Descriptions on the pole structures of $x_i$ dependent terms in ref.[2] were corrected, while the "determinant-ratio formulae" in [2] are correct.

The whole of the moduli dependent factor $E_K(\tau)$ is of a type of Schur polynomial. Its polynomial degree is $K-2$ because the degree of $Q_K(e_\delta)$ is $K$, and so its modular weight is $2(K-2)$ including contributions from all terms of $Q_K(e_\delta)$. Since it is known that the dimension of the modular space of the weight 2L is one for 2L = 4,6,8,10,14, the whole of this factor is proportional to $G_4(\tau), G_6(\tau), G_8(\tau), G_{10}(\tau)$, $G_{14}(\tau)$ for K = 4,5,6,7,9 respectively.

In this subsection, we saw that the Dolan-Godard generating function method with an auxiliary variable $v$ was quite efficient; this made our problem remarkably easier. Starting from the expansion form (2.43), we can obtain the coefficients $H_{N,M}(x_i)$ by utilizing the simplicity of the pole structures of Pe function (2.41). Once those are obtained, we can also derive the decomposition formula of $\prod_{i=1}^{N} S_\delta(x_i)$ by setting $v = \omega_\delta$. These two could be done almost separately. It is expected that the same method can be applied for higher genus cases. In higher genus, structures of singular part of genus g Pe function and structures of sigma function are not yet well known, and we can't do the similar calculations rigorously at present. Once these fundamental issues are clarified, the method will be directly generalized.

## 7) symmetrization

In eq.(6.18) of ref.[8], an addition formula is reported

$$\sum_{I_1,...I_N}^{g} \frac{\partial^N \ln \theta[\delta](0)}{\partial_{I_1}....\partial_{I_N}} \omega_{I_1}(z_1)....\omega_{I_N}(z_N) + \delta_{N,2}\omega(z_1,z_2) = +\sum_{\Gamma} (\prod_{l\in\Gamma} S_l) \qquad (2.49)$$

where $\Gamma$ is a connected oriented loop which passes through the points $z_1, z_2,...z_N$ once and for $l = (z_i, z_j) \in \Gamma : S_l = S_\delta(z_i, z_j)$.

In another word, the summation is constructed as follows. We prepare N vertices, 1,2,.....N and starting from 1, we choose the next vertex, say $i_2$ out of N-1 vertices. Then choose next vertex, $i_3$ out of the rest N-2 vertices, and so on. At the end, one vertex $i_N$ remains, and we make one product

$$S_\delta(z_1, z_{i_2})S_\delta(z_{i_2}, z_{i_3})S_\delta(z_{i_3}, z_{i_4})....S_\delta(z_{i_N}, z_1). \qquad (2.50)$$

The summation in (2.49) is all over possible permutations of $(N-1)!$ of such products. We do not divided by the factor $(N-1)!$, and we denote this summation, the right hand side of(2.49), as $<<$ $>>$ and call this "symmetrization" temporally.
The eq.(2.49) says that, after the symmetrization $<<$ $>>$, the (2.50)becomes theta constants except the case N=2. In eq.(2.49), we understand here that $\delta$ is the index of even theta function at genus g.
The special cases $N = 2$ and $N = 4$ of the formula (2.49) include two corollaries described in eq.(39) and eq.(40) in Fay's book[16].
If N is odd, the symmetrization gives zero. In the following, we assume that N is even and N > 2.

Let us calculate $<<$ $>>$ in (2.3) or (2.10) by borrowing the result in ref.[3]. The "symmetrization" is the same as S in ref.[3].

$$\prod_{i=1}^{N} S_\delta(z_i) = \frac{1}{(N-1)!} \sum_{K=0}^{\frac{N}{2}} H_{N,N-2K}(z_1, z_2,...z_N) \cdot P^{(2K-2)}(\omega_\delta) \qquad (2.10)$$

Here the variables of vertex inserting points $z_1, z_2,...z_N$ are used instead of $x_1, x_2,...x_N$.

It is shown in the Appendix B of ref.[3] that, except $N = 2$,

$$<< H_{N,N-2K}(z_1, z_2,...z_N) >> = 0 \qquad (2.51)$$

for the case that $N$ is even and $0 \le K < \dfrac{N}{2}$ [3]

Then, when symmetrized, only the term $K = \dfrac{N}{2}$ gives non-zero value if $N > 2$, and we have

$$<<\prod_{i=1}^{N} S_\delta(z_i)>> = \frac{1}{(N-1)!} <<H_{N,0}(z_1,z_2,...z_N)>> \cdot P^{(N-2)}(\omega_\delta)$$

$$= \frac{1}{(N-1)!}<<1>> \cdot P^{(N-2)}(\omega_\delta) \quad = \cdot P^{(N-2)}(\omega_\delta) \tag{2.52}$$

On the other hand, at genus 1, from eq. (2.49),

$$<<\prod_{i=1}^{N} S_\delta(z_i)>> = \frac{\partial^N \ln\theta_{\delta+1}(0)}{\partial x^N} \tag{2.53}$$

for three even theta functions $\theta_{\delta+1}$.

The equality $\dfrac{\partial^N \ln\theta[\delta](0)}{\partial x^N} = P^{(N-2)}(\omega_\delta)$ can be seen in genus 1 as follows.

From sigma function, we define the following three functions, often called co-sigma functions:

$$\sigma_\delta(x) = \exp(-\eta_\delta x)\frac{\sigma(x+\omega_\delta)}{\sigma(\omega_\delta)} \tag{2.54}$$

By using this notation, we can write (2.26) as

$$S_\delta(x) = \frac{\sigma_\delta(x)}{\sigma(x)} \tag{2.55}$$

On the other hand, there is a formula when $2\omega_1 = 1$

---

[3] If $N > 2$, $<<H_{N,N}(z_1,z_2,...z_N)>> = 0$. In the notations here, eq.(B23) in ref[3] may mean that the expansion of $v^{-n}H_n^S$ has only one term $\dfrac{(-1)^n}{(n-1)!}P^{(n-2)}(v)$ for $n > 2$, which means only $H_{n,0}^S$ is non-zero and for others are zero, $H_{n,m}^S = 0$, $m \le n$. In particular $H_{n,n}^S = 0$. For example, in (C.19) in ref[3], $H_{4,4}^S = 0$, because $H_{4,2}^S = 0$ and

$$[\chi_{NS}^{21}\chi_{NS}^{32}\chi_{NS}^{43}\chi_{NS}^{14}]^S = \frac{1}{6}<<\chi_{NS}^{21}\chi_{NS}^{32}\chi_{NS}^{43}\chi_{NS}^{14}>> = -\frac{\pi^4}{3}\theta_2(0,\tau)^4\theta_4(0,\tau)^4 = e_2^2 - \frac{1}{12}g_2 = \frac{1}{6}P^{(2)}(\omega_2)$$

as in (C.20), and cancels with the last constant term in (C.19).

$$\sigma_\delta(x) = \exp(\eta_1 x^2)\frac{\theta_{\delta+1}(x)}{\theta_{\delta+1}(0)} \qquad (2.56)$$

Then, differentiate the both side of the log of the following identity N times ( N>2, even )

$$\exp(-\eta_\delta x\ )\frac{\sigma(x+\omega_\delta)}{\sigma(\omega_\delta)} = \exp(\eta_1 x^2)\frac{\theta_{\delta+1}(x)}{\theta_{\delta+1}(0)} \qquad (2.57)$$

we have , for N>2,

$$\frac{\partial^N \ln\sigma(x+\omega_\delta)}{\partial x^N} = P^{(N-2)}(x+\omega_\delta) = \frac{\partial^N \ln\theta_{\delta+1}(x)}{\partial x^N} \qquad (2.58)$$

The equation (2.49) is the result of genus g, and all others are on genus 1. We will show later the genus g version of hyper elliptic case of (2.58). Eq.(2.49) suggests that, there will exist genus g- analog of the decomposing formula (2.10) written in Pe functions, which has more information than eq.(2.49), and only when it is symmetrized, such generalized equation reduces to (2.49 ). There will also exist all other formulas on genus g sigma function analogous to (2.54) - (2.58 ) for general surfaces.

**More on V**

The contents of this subsection are not used later.    Assume again that N is even.

The function $V_{N,M}(z_1, z_2,...z_N)$ defined as in the below of eq.(2.8) , which was obtained starting from another expression of Frobenius – Stickelberger formula (2.7), contains only the Pe function  and first derivative of Pe function. Everything is expressed without higher derivative terms. Eq.(2.8) also has simple coefficients $(e_\delta)^K$.

If, the derivatives of Pe function in (2.34)

$$\prod_{i=1}^N \frac{\sigma(x_i+v)}{\sigma(x_i)\sigma(v)} = \sum_{M=0}^N H_{N,N-M}(z_i)P^{(N-2-M)}(v) \qquad (P^{(-2)} \equiv 1 \text{ and } P^{(0)} \equiv P\ ) \qquad (2.34)$$

are expanded as

$$P^{(2n-2)}(v) = c_n P^n(v) + c_{n-1} P^{n-1}(v) + \ldots = \sum_{i=0}^n c_i P^i(v) \qquad (2.59)$$

Then it gives an expansion  by  $V_{N,N-M}$.   Therefore, expanding $Q_K(e_\delta)$ as a polynomial of $e_\delta$ in

$$\prod_{i=1}^N S_\delta(z_i - z_{i+1}) = \frac{1}{(N-1)!} \sum_{K=0}^{\frac{N}{2}} H_{N,N-2K}(z_1, z_2,...z_N) \cdot Q_K(e_\delta)$$

we may say that the coefficient of each degree of $e_\delta$ should be the same as $V_{N,N-2K}$.

When $Q_K(e_\delta)$ is expanded like (2.59), we write

$$Q_K(e_\delta) = \sum_{n=1}^{K} C_n(K)(e_\delta)^n \tag{2.60}$$

where n means the degree. Then

$$\prod_{i=1}^{N} S_\delta(z_i - z_{i+1}) = \frac{1}{(N-1)!} \sum_{K=0}^{\frac{N}{2}} H_{N,N-2K}(z_1, z_2, \ldots z_N) \cdot \sum_{n=1}^{K} C_n(K)(e_\delta)^n \tag{2.61}$$

Re-arranging the terms, we have

$$= \frac{1}{(N-1)!} \sum_{K=0}^{\frac{N}{2}} \sum_{M=K}^{\frac{N}{2}} H_{N,N-2M}(z_1, z_2, \ldots z_N) \cdot C_K(M)(e_\delta)^K \tag{2.62}$$

Note that the second summations begin at K, as well as the second index of H is N-2M. Then

$$V_{N,N-2K}(z_1, z_2, \ldots z_N) = \frac{1}{(N-1)!} \sum_{M=K}^{\frac{N}{2}} H_{N,N-2M}(z_1, z_2, \ldots z_N) \cdot C_K(M) \tag{2.63}$$

Therefore $V_{N,N-2K}(z_1, z_2, \ldots z_N)$ also has "single pole" structure in the variables $x_1, x_2, \ldots x_N$, although $\underset{x_N=0}{\text{Res}} V_{N,N-2K}(z_1, z_2, \ldots z_N)$ is not equal to $V_{N-1,N-2K-1}(z_1, z_2, \ldots z_N)$.

Calculating symmetrization, << >> as

$$<<V_{N,N-2K}(z_1, z_2, \ldots z_N)>> = \frac{1}{(N-1)!} \sum_{M=K}^{\frac{N}{2}} <<H_{N,N-2M}(z_1, z_2, \ldots z_N)>> \cdot C_K(M)$$

$$= \frac{1}{(N-1)!} <<H_{N,0}(z_1, z_2, \ldots z_N)>> \cdot C_K(\frac{N}{2}) \quad = \cdot C_K(\frac{N}{2}) \tag{2.64}$$

then

$$Q_{\frac{N}{2}}(\nu) = P^{(N-2)}(\nu) = \sum_{K=1}^{\frac{N}{2}} C_K(\frac{N}{2}) P^K(\nu) = \sum_{K=1}^{\frac{N}{2}} <<V_{N,N-2K}>> P^K(\nu) \tag{2.65}$$

This means an interesting feature: When the derivatives of Pe function $P^{(N-2)}(\nu)$ for even N is expanded by the Pe function itself, its coefficients are $<<V_{N,N-2K}>>$.

That is, $V_{N,N-2K}$, originally defined as a ratio of determinants in the description below (2.8) and are functions of, $z_1, z_2, ... z_N$, becomes such coefficients which are modular invariant constants written by Eisenstein series after summing up all (N-1)! terms of re-shuffled variables $z_1, z_2, ... z_N$.

Since $<<H_{N,N-2K}(z_1, z_2, ... z_N)>>$ is non-zero only if $N - 2K = 0$, we can write

$$P^{(N-2)}(v) = \sum_{K=1}^{\frac{N}{2}} \frac{1}{(N-1)!} <<H_{N,N-2K}>> P^{(2K-2)}(v) \qquad (2.66)$$

and

$$P^{(N-2)}(v) = \sum_{K=1}^{\frac{N}{2}} <<V_{N,N-2K}>> P^K(v) \qquad (2.67)$$

## 3. Genus $g > 1$
### 3-1

When considering the product of Szego kernels $S_\delta(z_1, z_2) S_\delta(z_2, z_3) ..... S_\delta(z_N, z_{N+1})$ with a constraint $z_{N+1} = z_1$ at genus g, from the argument in subsection 6.1, we may seek a function of the generalization of genus 1 case $\prod_{i=1}^{N} \frac{\sigma(x_i + v)}{\sigma(x_i)\sigma(v)}$ and expand it by genus g Pe function. Such a function should be equal to the product of Szego kernels if $v \in C^g$ equals to genus g half periods $\omega_\delta \in C^g$. Here, we restrict to the case that $\omega_\delta$ is non-singular and even, so that the theta functions in the Szego kernels are even.

A natural, perhaps almost unique candidate of such function is

$$A \cdot \prod_{i=1}^{N} \frac{\sigma(I(z_i - z_{i+1}) + v)}{E(z_i, z_{i+1})\sigma(v)} \qquad (3.1)$$

where $\sigma$ is the genus g sigma function, and $v \in C^g$ is an auxiliary variable vector. This is probably how genus g sigma function is included in general superstring amplitudes or in current algebras without imposing artificial assumptions.
A definition of sigma function in genus g is shown in the next subsection in hyper elliptic case, but there exists more general definition in mathematical literatures.

The factor $A$ has the following quadratic form

$$A = \exp\left(+\frac{1}{2}\sum_{i=1}^{N}\sum_{I,J=1}^{g} u_I^i \eta_{IJ} u_J^i\right) \tag{3.2}$$

where $u_I^i$ is the I-th component of the Abel map of $I(z_i - z_{i+1})$, that is, $u_I^i = \int_{z_i}^{z_{i+1}} \omega_I$.

This factor comes from the fact that the denominator of (3.1) includes prime form, not sigma function as was the case in genus 1. This factor A cancels the inverse of the quadratic factor in sigma function. Please do not confuse holomorphic one forms $\omega$ with half periods $\omega_\delta$. Usually the half periods are written as $\Omega_\delta$ but the symbol $\Omega$ is used as one of period matrices in this document.

As described above, in the following we cope with Riemann surfaces where the following equation is valid:

$$A \prod_{i=1}^{N} \frac{\sigma(I(z_i - z_{i+1}) + \omega_\delta)}{E(z_i, z_{i+1})\sigma(\omega_\delta)} = S_\delta(z_1, z_2)S_\delta(z_2, z_3)\ldots S_\delta(z_N, z_{N+1}) \tag{3.3}$$

with $z_{N+1} = z_1$ and non-singular even half periods $\omega_\delta$.

As for hyper elliptic cases, $\omega_\delta$ have explicit and simple form as in (3.23) and a derivation of (3.3) is explained in the next subsection. It is expected that eq. (3.3) is valid for more general curves.

The prime form in (3.1) has holomorphic one forms. By the condition $z_{N+1} = z_1$, those have the form $h^2(z_1)h^2(z_2)\ldots h^2(z_N)$ where

$$h(z_i) = \sqrt{\sum_I \partial_I \vartheta[\lambda](0)\omega_I(z_i)} \tag{3.4}$$

and $\lambda$ is a suffix of one odd theta functions used to define prime form. Since for each fixed value of $I_1, I_2, \ldots I_N$ ( each of $I_i$ varies from 1 to g) in $\omega_{I_1}(z_1)\omega_{I_2}(z_2)\ldots \omega_{I_N}(z_N)$ the function can be expanded by Pe function, and so the function (3.1) will have the following expansion form:

$$A \cdot \prod_{i=1}^{N} \frac{\sigma(I(z_i - z_{i+1}) + v)}{E(z_i, z_{i+1})\sigma(v)} = \sum_{I_1, I_2, \ldots I_N = 1 (perm)}^{g} [H_0^N(z_i)P_{I_1 I_2 \ldots I_N}(v) + H_{I_1}^N(z_i)P_{I_2 \ldots I_N}(v) + H_{I_1 I_2}^N(z_i)P_{I_3 \ldots I_N}(v) + \ldots$$

$$......+ H^{N}_{I_1I_2...I_{N-2}}(z_i)P_{I_{N-1}I_N}(\nu) + H^{N}_{I_1I_2...I_{N-1}}(z_i)P_{I_N}(\nu) + H^{N}_{I_1I_2...I_N}(\nu)]\omega_{I_1}(z_1)\omega_{I_2}(z_2)......\omega_{I_N}(z_N)$$

$$= \sum_{I_1,I_2,...I_N=1 \,(perm)}^{g} \sum_{M=0}^{N} [\sum H^{N}_{I_1I_2...I_M}(z_i)P_{I_{M+1}...I_N}(\nu)]\omega_{I_1}(z_1)\omega_{I_2}(z_2)......\omega_{I_N}(z_N)$$

(3.5)

The genus g Pe functions and its derivatives are defined as

$$P_{JK} = -\frac{\partial^2}{\partial u_J \partial u_K}\ln\sigma(u) \qquad P_{I_1I_2...I_N} = \frac{\partial^{N-2}}{\partial u_{I_3}\partial u_{I_4}....\partial u_{I_{N1}}}P_{I_1I_2} \qquad (3.6)$$

In (3.5), a summation "perm" is introduced so that N number of indices $I_1, I_2,... I_N$ of $H$ and $P$ should attach to the vertex inserting points $z_1, z_2,... z_N$ equally.

The coefficient $.H^{N}_{I_1I_2...I_{N-1}}(z_i)$ in front of $.P_{I_N}(\nu)$ will be zero but we do not argue it here.

Since we have adopted a concrete function $A\prod_{i=1}^{N}\frac{\sigma(I(z_i - z_{i+1}) + \nu)}{E(z_i, z_{i+1})\sigma(\nu)}$, the coefficients $H$ are determined by this expansion. This is equivalent to the generating function method of genus g case. The $H$ are the very modular invariant functions of vertex inserting points which remain unaffected in the process of spin structure sum, and will be included in the final form of general g-loop N-point superstring amplitudes.

It is desirable to have explicit form of $H$. This process may be done, as in the case of genus 1, by multiplying a function of $\nu$ on the both sides of the expansion (3.5) so that the both sides do not have singular part of $\nu$, and then differentiate with respect to $\nu_1, \nu_2,....\nu_g$ which are components of $\nu$.

It is known that the genus g sigma function has a leading term so called Schur-Weierstrass polynomial $S(\nu)$, where $\nu_1, \nu_2,....\nu_g$ are components of $\nu$, as follows.

$$\sigma(v) = S(v) + \textit{higher terms of } v_i \tag{3.7}$$

Examples of $S(v)$ are

$$g = 1 \quad S(v_1) = v_1$$

$$g = 2 \quad S(v_1, v_2) = v_1 - \frac{1}{3}v_2^3 \tag{3.8}$$

$$g = 3 \quad S(v_1, v_2, v_3) = \frac{1}{45}v_3^6 - \frac{1}{3}v_2 v_3^3 - v_2^2 + v_1 v_3 \tag{3.9}$$

It would be good if the both side of (3.5) have no singular terms of $v$ after a power of Schur-Weierstrass polynomial is multiplied, but seems it is not so easy to prove this fact. Here we *assume* as follows:

There exists a function $X(v) \in C^1$, $v \in C^g$ such that the product $[X(v)]^N P_{I_1 I_2 \ldots I_N}(v)$ has no singular terms in the expansion:

$$[X(v)]^N P_{I_1 I_2 \ldots I_N}(v) = const + \textit{higher de}gree \textit{ terms of } v_1, v_2, \ldots v_g \quad . \tag{3.10}$$

In genus 1, $X(v) = v$ , $\dfrac{\sigma(v)}{X(v)} = \prod_{m,n}(1 - \dfrac{v}{\Omega_{m,n}})\exp[\dfrac{v}{\Omega_{m,n}} + \dfrac{v^2}{2\Omega_{m,n}^2}]$, and $X(v)$ gives singular part of derivatives of Pe function.

If (3.10) is assumed, as we saw for the g=1 case in subsection 6.2, the coefficients $H^N_{I_1 I_2 \ldots I_M}$ are basically obtained by differentiating appropriate times of

$$[X(v)]^N A \prod_{i=1}^{N} \frac{\sigma(I(z_i - z_{i+1}) + v)}{E(z_i, z_{i+1})\sigma(v)}$$ with respect to the variables $v_1, v_2, \ldots v_g$ and then setting all of them equal to 0, using eq.(3.5). We write

$$\left[\frac{X(v)}{\sigma(v)}\right]^N A \prod_{i=1}^{N} \frac{\sigma(I(z_i - z_{i+1}) + v)}{E(z_i, z_{i+1})} = A\exp\left[\ln \sigma(I(z_i - z_{i+1}) + v) - \ln E(z_i, z_{i+1}) - N\ln\left[\frac{\sigma(v)}{X(v)}\right]\right]$$

$$= \sum_{I_1, I_2, \ldots I_N = 1}^{g} \exp\left[\ln \sigma(I(z_i - z_{i+1}) + v) - \ln \overline{E}(z_i, z_{i+1}) - N\ln\frac{\sigma(v)}{X(v)}\right]\omega_{I_1}(z_1)\ldots\omega_{I_N}(z_N)$$

$$\tag{3.11}$$

where

$$\overline{E}(z_i, z_{i+1}) \equiv \frac{\vartheta[\lambda](I(z_1 - z_2)) \cdots \vartheta[\lambda](I(z_N - z_1)) \cdot}{\partial_{I_1} \vartheta[\lambda](0) \cdot \partial_{I_2} \vartheta[\lambda](0) \cdot \ldots \partial_{I_N} \vartheta[\lambda](0)} \cdot \frac{1}{\ln A} \tag{3.12}$$

for a fixed set of $I_1, I_2, \ldots I_N$. Then, define $F^g(v)$ as

$$F^g(v) = \exp\left[\ln \sigma(I(z_i - z_{i+1}) + v) - \ln \overline{E}(z_i, z_{i+1}) - N \ln \frac{\sigma(v)}{X(v)}\right], \tag{3.13}$$

and expand this in terms of $v_1, v_2, \ldots v_g$ to obtain explicit form of $H^N_{I_1 I_2 \ldots I_M}$.

If there are $a_1$ times of 1 in $I_1, I_2, \ldots I_N$, $a_2$ times of 2 in $I_1, I_2, \ldots I_N$, ..., $a_N$ times of N in $I_1, I_2, \ldots I_N$, we have

$$H^N_{I_1 I_2 \ldots I_M} = const \frac{\partial^M}{\partial v_1^{a_1} \partial v_2^{a_2} \ldots \partial v_N^{a_N}} F^g(v)\bigg|_{v=0} = const \frac{\partial^M}{\partial v_{I_1} \partial v_{I_2} \ldots \partial v_{I_M}} F^g(v)\bigg|_{v=0}$$

(3.14)

where $a_1 + a_2 + \ldots + a_N = M$. (3.15)

Since in general

$$P_{I_1 I_2 \ldots I_M}(\omega_\delta) = 0 \quad \text{if M is odd and}$$

$P_{I_1 I_2 \ldots I_M}(v)$ can be expressed as a polynomial of $P_{IJ}(v)$ if M is even, (3.16)

the following identity hold by (3.3) :

$$\prod_{i=1}^{N} S_\delta(z_i, z_{i+1}) = \sum_{I_1, I_2, \ldots I_N = 1 \, (perm)}^{g} \sum_{K=0}^{\left[\frac{N}{2}\right]} [\sum H^N_{I_1 I_2 \ldots I_{N-2K}}(z_i) P_{I_{N-2K+1} \ldots I_N}(\omega_\delta)] \omega_{I_1}(z_1) \omega_{I_2}(z_2) \ldots \omega_{I_N}(z_N)$$

(3.17)

and the spin structure dependence of the product $\prod_{i=1}^{N} S_\delta(z_i, z_{i+1})$ is totally expressed by one kind of constants, $P_{IJ}(\omega_\delta)$, Pe function values at the non singular even half periods. This is a natural extension of the case of genus 1.

The statements in (3.16) are valid at least for hyper elliptic curves. It is expected that these holds for more general cases too. If it is not the case, we use (3.5) with $v = \omega_\delta$.

Since $v_1, v_2, \ldots v_g$ are all set equal to zero after the differentiations, the details of the structure of $X(v)$ affects only numerical constant factors in (3.14). The constants in front of the right hand side of (3.14) are determined once $X(v)$ is clarified.

The examples of explicit form of $H^N_{I_1 I_2 ... I_M}$ will be as follows, up to over all numerical constants.

$$H^N_0 = const \tag{3.18}$$

$$H^N_{I_1} = \frac{\partial}{\partial v_{I_1}} F^g(v)\Big|_{v=0} = \sum_{i=1}^{N} \frac{\partial}{\partial v_{I_1}} \ln \sigma(I(z_i - z_{i+1})) \tag{3.19}$$

$$H^N_{I_1 I_2} = \frac{1}{2} \frac{\partial^2}{\partial v_{I_1} \partial v_{I_2}} F^g(v)\Big|_{v=0}$$

$$= \frac{1}{2}[\sum_{i=1}^{N} \frac{\partial}{\partial v_{I_1}} \ln \sigma(I(z_i - z_{i+1}))][\sum_{i=1}^{N} \frac{\partial}{\partial v_{I_2}} \ln \sigma(I(z_i - z_{i+1}))] - \frac{1}{2} \sum_{i=1}^{N} P_{I_1 I_2}(I(z_i - z_{i+1}))$$

$$\tag{3.20}$$

We assumed that $\ln \frac{\sigma(v)}{X(v)}$ does not contribute to $H^N_{I_1 I_2 ... I_M}$ for small values of M as in the case of genus 1. At genus 1, the factor $\ln \frac{\sigma(v)}{X(v)}$ can be replaced with Eisenstein series. It seems that the genus g version of the formula of (2.38) is not known in general case, but probably it exists, and $\ln \frac{\sigma(v)}{X(v)}$ will be replaced with genus g modular forms and give $z_i$ independent contribution, on the premise of setting $v_1, v_2, ...v_g \to 0$ after the differentiations. $\tag{3.21}$

The factor $\prod_{i=1}^{N} \sigma(v)$ in the denominator of the original function $A \cdot \prod_{i=1}^{N} \frac{\sigma(I(z_i - z_{i+1}) + v)}{E(z_i, z_{i+1})\sigma(v)}$ is obviously independent of $z_i$. This factor is in the last term of $F^g(v)$, (3.13), and if this contributes to modular invariant function $H^N_{I_1 I_2 ... I_M}$ by differentiations of $F^g(v)$, then its result will be inevitably constant modular forms of genus g, as in genus 1. If this factor contributes to the form of $H^N_{I_1 I_2 ... I_M}$ for $N \leq 3$, it may contradict the fact in genus 1 case when pinching the Riemann surface to $N$ tori. Also, after setting $v_1, v_2, \ldots v_g \to 0$ after differentiations, an over-all factor $\dfrac{\sigma(I(z_i - z_{i+1}))}{\bar{E}(z_i, z_{i+1})}$ appears in the results. This does not have poles of $z_{i+1} - z_i$ and will be constants, but further considerations together with $\ln \dfrac{\sigma(v)}{X(v)}$ should be needed.

The permutation summation $\sum_{(perm)}$ in (3.5) and (3.17) means that we also have to consider $H^N_{I_2}, H^N_{I_3}, \ldots H^N_{I_N}$ in (3.19), and all combinations of $I_j, I_k$ in (3.20).

In genus 1 and 2, four-point superstring amplitudes give non-zero result only from $H^N_0$. All other $H^N_{I_1 I_2 \ldots I_M}$ are for higher point amplitudes. When we calculate symmetrization of (3.17), the result will be (2.49).

In this section, we made two substantial assumptions to obtain explicit form of $H^N_{I_1 I_2 \ldots I_M}$, due to the difficulty of investigating detailed structure of sigma function and structures of singular part of derivatives of Pe function on the $v$ side. One is that the function (3.3) can be expanded as $\sum_{M=0}^{N} H^N_{I_1 I_2 \ldots I_M}(z_i) P_{I_{M+1} \ldots I_N}(v)$ in (3.5) where summation is only up to N, and Pe function has simple singular structure as of (3.10). If the structure of singular part of $P_{I_1 I_2 \ldots I_N}$ is clarified, many of things will be clarified. The other assumption is (3.21).

By these, as was the case in genus 1, the functions of vertex inserting points $z_i$ come only from differentiations of sigma function on the exponential factor $F^g(v)$, (3.13). As a consequence, polynomial forms of differentials of $\ln \sigma(I(z_i - z_{i+1}))$ are quite similar as those in genus 1, the difference is that in higher genus there are g numbers of variables to differentiate instead of one variable.

This is a nice simplicity since the product of fermion correlation functions in genus g has the same pole structure as that in genus 1 for $z_i - z_{i+1}$ side, simultaneous single pole structure in the variables $z_i - z_{i+1}$, and the right hand side of (3.17) should have such pole structures. The forms of the functions are basically determined, although numerical constant factors and contributions from the genus g modular forms are not determined. Strictly speaking, the possibility of expansion (3.5) with N terms on the side of $v$ needs to be argued mathematically in higher genus. Considering the poles on the side of $z_i - z_{i+1}$, such expansion as well as the assumptions we made look natural. Also, eq.(2.49), which was obtained long time ago, suggests the validity of the expansion form of (3.5) as its generalization.

These poles of $z_i - z_{i+1}$ will be particle poles of g loop N point superstring amplitudes in the same manner as in genus 1.

If rigorous formulae including numerical factors about the decomposition identities are obtained, the results of H will be used to describe Kac-Moody currents in a closed form of genus g.

### 3-2 Hyper elliptic case

It is instructive to consider hyper elliptic cases because we can see explicit forms of half periods corresponding to non-singular even spin structures as integrals given in (3.23) below and can see how arguments in the previous subsection work in concrete calculations. We also would like to point out that in hyper elliptic case there is a method to calculate explicitly the key constants $P_{IJ}(\omega_\delta)$ helped by an elegant classical theory related to Jacobi inversion problem[2][12]. As described in the previous subsection, spin structure dependence of amplitudes are entirely determined only by this one kind of constants $P_{IJ}(\omega_\delta)$ for any g and N at least for hyper elliptic cases.

If $H^N_{I_1 I_2 \cdots I_M}$ are actually given as in (3.15) type formula, all tools to calculate the spin structure sum of $S_\delta(z_1, z_2) S_\delta(z_2, z_3) \cdots S_\delta(z_N, z_{N+1})$ with a constraint $z_{N+1} = z_1$ for arbitrary g loop N points can be prepared as follows, along a scenario guessed in ref.[2].

Consider the branch points of the curve, $e_1, e_2, \ldots e_{2g+2}$, and fix the value of $e_{2g+2}$ at $\infty$. As long as non-singular even spin structures are concerned, there is one to one correspondence between one spin structure and one choice of $g$ number of branch points, say $f_1, f_2, \ldots f_g$, out of $2g+1$ points $e_1, e_2, \ldots e_{2g+1}$. These can be seen as follows. First, there are $\binom{2g+1}{g}$ possible choices, and this number is equal to $\frac{1}{2}\binom{2g+2}{g+1}$, which is the way of grouping $2g+2$ branch points into two parts, and is equal to the number of non-singular even spin structures.

Consider the abelian image of each of branch points as $U_j \equiv \int_\infty^{(e_j,0)} \omega \in C^g$ .
This integral can be done explicitly. For all of holomorphic one forms $\omega_i$, $i = 1,2,...g$ the results of the integrals have a form

$$U^i_j = \int_\infty^{e_j} \omega_i \equiv E^i_j + \Omega \overline{E}^i_j \tag{3.22}$$

where both of $E_j$ and $\overline{E}_j$ for each j are dimension g vectors whose all elements are zero or 1/2. Obviously, if $j = 2g+2$, all components of $E_{2g+2}$ and $\overline{E}_{2g+2}$ are zero. See the concrete example of g=2 in the Appendix C.

The $U_j$ themselves are half periods, and the summations over the following g number of branch points chosen out of 2g+1 number are also half periods:

$$\omega_\delta \equiv \sum_{m=1}^{g} \int_\infty^{(f_m,0)} \omega \in C^g \tag{3.23}$$

Delta denotes one choice of g branch points. This is the explicit definition of half periods $\omega_\delta$ used in this document, if the curve is hyper elliptic, and this corresponds to non singular even spin structures as we will see below.

It is known that the vector of Riemann constants has also this type of summation:

$$\Delta^i = -\sum_{k=1}^{g} \int_\infty^{e_{2k}} \omega_i = -\sum_{k=1}^{g} U^i_{2k} \quad . \tag{3.24}$$

We define the g component vector indices $\Delta^a, \Delta^b$ corresponding to the Riemann constant by the following equation:

$$\Delta = \Delta^a + \Omega \Delta^b \tag{3.25}$$

The genus g hyper elliptic sigma function is defined using a theta function whose indices are the Riemann constant ,

$$\sigma(u) = c \exp\left(-\frac{1}{2} \sum_{I,J=1}^{g} u_I \eta_{IJ} u_J\right) \theta\begin{bmatrix} \Delta^a \\ \Delta^b \end{bmatrix}(u) \ . \tag{3.26}$$

Here the Jacobi theta function is defined in a standard notation:

$$\theta\begin{bmatrix} a \\ b \end{bmatrix}(u, \Omega) = \sum_{n \in Z^g} \exp\{i\pi(n+a)^t \Omega(n+a) + 2\pi i(n+a)(u+b)\} \tag{3.27}$$

$u \in C^g$

The theta function is called odd or even depending on whether the 4ab is even or odd. This can also be written as

$$\theta\begin{bmatrix} a \\ b \end{bmatrix}(u, \Omega) = \theta(u+b+\Omega a, \Omega) \exp\{i\pi a^t \Omega a + 2\pi i a(u+b)\} \tag{3.28}$$

where $\theta(u, \Omega)$ is the standard theta function with zero index.

Suppose that half periods are included in the variables of sigma function:

$$u = \bar{u} + \omega_\delta . \tag{3.29}$$

There are some situations in which the exp factor in (3.28), $\exp\{i\pi a^t \Omega a + 2\pi i a(u+b)\}$, can be disregarded. The left hand side of eq.(3.3), $A \cdot \prod_{i=1}^{N} \frac{\sigma(I(z_i - z_{i+1}) + v)}{E(z_i, z_{i+1})\sigma(v)}$, is one of such examples when $v$ is set equal to $\omega_\delta$. This is a ratio of two sigma functions, and considering the condition $z_{N+1} = z_1$, the exponential factor $\exp\{i\pi a^t \Omega a + 2\pi i a(u+b)\}$ becomes 1.

In such cases, by seeing the form of variables $\theta(u+b+\Omega a, \Omega)$ in eq.(3.28), the $\omega_\delta$ only has the effect to shift the index of the theta function from $\Delta$ to $\omega_\delta + \Delta$. And a nice fact is that it is known that the theta functions which have indices $\omega_\delta + \Delta$ are even. Actually this is the definition of the suffix $\delta$ which denotes the even theta functions.

This is how the left hand side changes to the product of fermion correlation functions which includes even theta functions.

There is one more situation where the factor $\exp\{i\pi a^t \Omega a + 2\pi i a(u+b)\}$ can be disregarded. The differentiation of $\ln \sigma(u)$ with respect to the I-th component of $u$ will give such a situation. Even if the differentiation is only one time, due to the condition $z_{N+1} = z_1$, we can disregard that factor.

Then, on hyper elliptic curves, we have

$$\frac{\partial^N \ln \theta[\delta](u)}{\partial_{I_1}....\partial_{I_N}} = P_{I_1 I_2...I_N}(u+\omega_\delta) \tag{3.30}$$

for N>2, which is a generalization of the genus 1 formula, eq. (2.58).

Let us see the right hand side of the decomposition formula (3.17).
we saw that all information about one fixed spin structure is included in $P_{IJ}(\omega_\delta)$. In $g=1$, this has only one elements for each spin structure, $P(\omega_\delta)$, and $P(\omega_\delta)$ is equal to the branch point $e_\delta$ itself. In genus $g>1$, $P_{IJ}(\omega_\delta)$ which has $\frac{g(g+1)}{2}$ elements as a symmetric matrix of the indices $I,J$, can also be represented by the branch points themselves, helped by the method of solving so called Jacobi inversion problem [11][12][14], as follows.

First, when one of the index is equal to $g$, that is, $P_{Ig}$, are shown to be equal to elementary symmetric polynomials of the chosen g branch points $f_1, f_2,.....f_g$, except the overall sign +1, -1.

The $P_{1g}$ has the highest degree of fundamental symmetric polynomials, $g$, and for $P_{2g}$, $P_{3g}$, ..... $P_{gg}$, the degree is decreasing as $g-1$, $g-2$, ..... 1.

The rest $\frac{g(g+1)}{2} - g = \frac{g(g-1)}{2}$ numbers of elements of $P_{IJ}$ are determined as follows. Expand the curve $y^2 = 4\prod_{k=1}^{2g+1}(x-e_k)$ as

$$4\prod_{k=1}^{2g+1}(x-e_k) = 4x^{2g+1} + \mu_1 x^{2g} + \mu_2 x^{2g-1} + .... + \mu_{2g+1} \tag{3.31}$$

where $\mu_1, \mu_2,.....\mu_{2g+1}$ are elementary symmetric polynomials of $e_1, e_2,.....e_{2g+1}$.

Construct a concrete polynomial $F(x,z)$ as:

$$F(x,z) = \sum_{i=0}^{g} x^i z^i (\mu_{2g-2i}(x+z) + 2\mu_{2g-2i+1})$$

$$= \{(x+z) + 2\mu_1\}x^g z^g + \{\mu_2(x+z) + 2\mu_3\}x^{g-1}z^{g-1} + ... + \mu_{2g}(x+z) + 2\mu_{2g+1} \tag{3.32}$$

Then, the following equations hold for any choice of two branch points $x_r$, $x_s$ from the chosen $g$ number of points $f_1, f_2, ..... f_g$ :

$$\sum_{I=1}^{g} \sum_{J=1}^{g} P_{IJ}(\omega_\delta) x_r^{I-1} x_s^{J-1} = \frac{F(x_r, x_s)}{(x_r - x_s)^2} \qquad (3.33)$$

There are $\binom{g}{2} = \frac{g(g-1)}{2}$ number of such equations, and it matches the necessary number of equations to determine all elements $P_{IJ}(\omega_\delta)$ by branch points.

In superstring theories, if only hyper elliptic curve is concerned, the string measure is constructed by branch points[10]. Then summing over spin structure is an algebraic calculation of branch points as $\sum (string\ measure) \cdot P_{IJ}(\omega_\delta)$, irrespective of functions of vertex inserting points H, as was the case in genus 1. The result of algebraic calculation will be represented by polynomials of symmetric functions of branch points, which will be expressed by theta constants and be modular forms at the genus g. As in genus 1, the modular forms will appear from two parts: One is here, and the other is from $\ln \frac{\sigma(v)}{X(v)}$ in eq.(3.13).

To do all calculations of superstring amplitudes this is not the whole of the story even in the hyper elliptic case. We have to consider connected parts of the contractions, as has been pointed out in ref.[9] and their related articles. These investigations are beyond the scope of this document. The Dolan-Goddard generating function method makes it possible to consider obtaining the forms of $H^N_{I_1 I_2 ... I_M}$ and moduli constant dependent parts almost separately for higher genus. A lot of arguments are made on the latter (For example [9][17]-[22]) including superstring measures. To solve the former may be more straightforward once structures of genus g sigma function and Pe functions are clarified, and it may have applications beyond string theories.

**Acknowledgements**
The author thanks Prof. A.Morozov for his kind help in the process of submitting ref.[2] and this document to arXiv.
( This document will be submitted to arXiv only.)

## Appendix A    Proof of (2.3)

By the condition $\sum_{i=1}^{N} x_i = 0$, the right hand side of (2.2) becomes zero because $\sigma(x_1 + x_2 + \cdots x_N) = 0$, that is,

$$\det\nolimits_{NxN}[1, P, P^{(1)}, P^{(2)}, P^{(3)}, P^{(N-2)}](x_1, x_2, \ldots x_N) = 0 \tag{A.1}$$

Therefore, for $v = x_1, x_2, \ldots x_N$, there exist $a_0, a_1, \ldots a_{N-2}$ which satisfy

$$1 - a_0 P(v) - a_1 P^{(1)}(v) - a_2 P^{(2)}(v) - a_3 P^{(3)}(v) \cdots - a_{N-2} P^{(N-2)}(v) = 0 \tag{A.2}$$

Minus signs in front of $a_0, a_1, \ldots a_{N-2}$ are only for notational convention.

This can solved for $a_0, a_1, \ldots a_{N-2}$, by Cramer's formula. There are $N$ linear equations for $N-1$ number of variables; this is a result of redundant relation $\sum_{i=1}^{N} x_i = 0$. It is possible to choose any of $N-1$ equations or variables to solve. Choose the variables $x_1, \ldots x_{N-1}$, the equations are

$$mat_{(N-1)x(N-1)}[P, P^{(1)}, P^{(2)}, P^{(3)}, \cdots P^{(N-2)}](x_1, x_2, \ldots x_{N-1}) * (a_0, a_1, a_2, \cdots a_{N-2})^T$$

$$= (1,1,\ldots 1)^T \tag{A.3}$$

Then we have

$$a_i = \frac{\det_{(N-1)x(N-1)}[P, P^{(1)}, P^{(2)}, P^{(3)} \ldots P^{(N-2)}]((i+1)th \to 1 \,;\, x_1, x_2, \ldots x_{N-1})}{\det_{(N-1)x(N-1)}[P, P^{(1)}, P^{(2)}, P^{(3)} \ldots P^{(N-2)}](x_1, x_2, \ldots x_{N-1})} \tag{A.4}$$

Now we consider polynomials $f(x)$ and $h(x)$ defined as follows. For a while, we assume $N$ is even.

$$f(P(v)) \equiv \{h(P(v))\}^2 - \{a_1 P^{(1)}(v) + a_3 P^{(3)}(v) + a_5 P^{(5)}(v) \cdots + a_{N-3} P^{(N-3)}(v)\}^2 \tag{A.5}$$

$$h(P(v)) \equiv 1 - a_0 P(v) - a_2 P^{(2)}(v) - a_4 P^{(4)}(v) \cdots - a_{N-2} P^{(N-2)}(v) \tag{A.6}$$

That is, $f(x)$ is a polynomial of Pe function of a form $\{\sum P^{(EVEN)}\}^2 - \{\sum P^{(ODD)}\}^2$, and $h(x)$ is its even derivatives part. The $f(x)$ is a degree $N$ polynomial because it can be written as :

$$f(P(v)) = h(P(v))^2 - \{P^{(1)}(v)\}^2 [polynomial\ of\ P(v)] \tag{A.7}$$

and the degree of $P^{(N-2)}$ is $\dfrac{N}{2}$.

On the other hand, if we factorize the definition of $f(x)$ in (A.5) as $\{\sum P^{(EVEN)} + \sum P^{(ODD)}\}\{\sum P^{(EVEN)} - \sum P^{(ODD)}\}$ , then we can see that , the equation $f(x)=0$ has $N$ number of solutions at $x = P(x_1), P(x_2),....P(x_N)$ by (A.2). Therefore, $f(x)$ can be written in a different way :

$$f(x) = c \cdot a_{N-2}^2 (x - P(x_1))(x - P(x_2)).....(x - P(x_N)) \tag{A.8}$$

The constant c is strictly determined so that the highest degree of P should be equal to that of (A.5). Since $P^{(N-2)}(\nu) = (N-1)! P^{\frac{N}{2}}(\nu) + ......$, c is determined as $c = \{(N-1)!\}^2$.

If we compare $f(x)$ with the form of $\prod_{i=1}^{N} S_\delta(x_i)$, because of the relationship $S_\delta(\nu)^2 = P(\nu) - e_\delta$ , the latter is found to be proportional to the square root of $f(x)$ with putting $x = e_\delta$. That is,

$$\prod_{i=1}^{N} S_\delta(x_i) = \dfrac{[f(e_\delta)]^{\frac{1}{2}}}{c^{1/2} \cdot a_{N-2}} \tag{A.9}$$

If we note $P(\omega_\delta) = e_\delta$ then $f(e_\delta) = f(P(\omega_\delta))$ , but since $P^{(1)}(\omega_\delta) = 0$ , we find from (A.8) that

$$f(e_\delta) = f(P(\omega_\delta)) = h(P(\omega_\delta))^2 - \{P^{(1)}(\omega_\delta)\}^2 [polynomial\ of\ P(\omega_\delta)] = h(P(\omega_\delta))^2 \tag{A.10}$$

Therefore the square root in (A.9) disappears, only leaving $h(e_\delta)$ :

$$\prod_{i=1}^{N} S_\delta(x_i) = \dfrac{[f(e_\delta)]^{\frac{1}{2}}}{(N-1)! \cdot a_{N-2}} = \dfrac{[f(P(\omega_\delta))]^{\frac{1}{2}}}{(N-1)! a_{N-2}} = \dfrac{h(e_\delta)}{(N-1)! a_{N-2}}$$

$$= \dfrac{1}{(N-1)!} \dfrac{1 - a_0 P(\omega_\delta) - a_2 P^{(2)}(\omega_\delta) - a_4 P^{(4)}(\omega_\delta) \cdots - a_{N-2} P^{(N-2)}(\omega_\delta)}{a_{N-2}} \tag{A.11}$$

Then, all of $P^{(EVEN)}(\omega_\delta)$ terms are represented by the polynomials of $e_\delta$, and the $x_i$ dependent terms are represented by N factors $\frac{1}{a_{N-2}}$, $-\frac{a_0}{a_{N-2}}$, $-\frac{a_2}{a_{N-2}}$, $-\frac{a_4}{a_{N-2}}$, ....$-\frac{a_{N-4}}{a_{N-2}}$, $-1$. These are the definition of $H_{N,N}$, $H_{N,N-2}$, ... $H_{N,0}$, and, using the solutions for $a_i$ in (A.4). We have to be careful that there is an over-all sign ambiguity when we calculate $\sqrt{(....)^2}$. We determined this so that the formula matches with $S_\delta^2(x) = P(x) - e_\delta$ when N=2. Since

$$S_\delta(x_1)S_\delta(x_2) = -S_\delta^2(x_1) = -P(x) + e_\delta = H_{2,2} + H_{2,0}e_\delta, \qquad (A.12)$$

we choose the overall sign by defining

$$H_{N,N} \equiv (-1) \frac{\det_{(N-1)x(N-1)}[P, P^{(1)}, P^{(2)}, P^{(3)}, ...... P^{(N-2)}](x_1, x_2,...x_{N-1})}{\det_{(N-1)x(N-1)}[P, P^{(1)}, P^{(2)}, P^{(3)}, ...... P^{(N-2)}]((N-1)th \to 1; x_1, x_2,...x_{N-1})}$$

(2.4)

$$H_{N,N-2K} \equiv + \frac{\det_{(N-1)x(N-1)}[P, P^{(1)}, P^{(2)}, P^{(3)}, ...... P^{(N-2)}]((2K-1)th \to 1; x_1, x_2,...x_{N-1})}{\det_{(N-1)x(N-1)}[P, P^{(1)}, P^{(2)}, P^{(3)}, ...... P^{(N-2)}]((N-1)th \to 1; x_1, x_2,...x_{N-1})}$$

$$(K > 0) \qquad (2.5)$$

$$H_{N,0} \equiv +1$$

to express the eq.(A.11) as

$$\prod_{i=1}^{N} S_\delta(x_i) = \frac{1}{(N-1)!} \sum_{K=0}^{\left[\frac{N}{2}\right]} H_{N,N-2K}(x_1, x_2,...x_N) \cdot P^{(2K-2)}(\omega_\delta) \qquad (2.3)$$

If N is odd, then we re-define the polynomial of (A.5) and (A.6) as

$$f(P(v)) \equiv \{h(P(v))\}^2 - \{a_1 P^{(1)}(v) + a_3 P^{(3)}(v) + a_5 P^{(5)}(v) \cdots + a_{N-3} P^{(N-2)}(v)\}^2 \qquad (A.13)$$

$$h(P(v)) \equiv 1 - a_0 P(v) - a_2 P^{(2)}(v) - a_4 P^{(4)}(v) \cdots - a_{N-2} P^{(N-3)}(v) \qquad (A.14)$$

Here the highest degree term $P^{(N-2)}(v)$ is included in the odd derivative part. Then, repeating the same argument, the last term of the numerator of (A.11) becomes

$-a_{N-2}P^{(N-3)}(\omega_\delta)$, instead of $-a_{N-2}P^{(N-2)}(\omega_\delta)$.

This means that we do not have to change anything in (2.4) and (2.5), as long as the summation in (2.3) is until $\left[\dfrac{N}{2}\right]$.

The proof of (2.8) as well as obtaining the form of $V_{N,M}(x_1, x_2, x_3...x_N)$ can be done in the similar way. For this, we need to change the definition of $f(x)$ and $h(x)$ in (A.5) (A.6) such that the derivatives of Pe function are not included but the monomials of Pe and $P^{(1)}$ are included.

As is shown in chapter 2 and 3 of this document, due to the simplicity of pole structures of derivatives of Pe function, it is always more convenient to argue on $H_{N,M}$ rather than $V_{N,M}$.

## Appendix B  POLES

In the following, $y_1, y_2, \cdots y_n$ are $n$ complex numbers.

Define the determinant $G(1, y_1, y_2, \cdots y_n)$ where the 1st degree of the variable is absent since this represents the poles of Pe function :

$$G(1, y_1, y_2, \cdots y_n) \equiv \begin{vmatrix} 1 & y_1^2 & y_1^3 & \cdots & \cdots & y_1^n \\ 1 & y_2^2 & y_2^3 & \cdots & \cdots & y_2^n \\ \vdots & \vdots & \vdots & & & \vdots \\ \vdots & \vdots & \vdots & & & \vdots \\ 1 & y_n^2 & y_n^3 & \cdots & \cdots & y_n^n \end{vmatrix} \quad (B.1)$$

Define $L(1, k, y_1, y_2, \cdots y_n)$ as follows :

$$L(1, k, y_1, y_2, \cdots y_n) \equiv \begin{vmatrix} 1 & y_1^2 & y_1^3 & \cdots & y_1^{k-1} & y_1^{k+1} & \cdots & y_1^{n+1} \\ 1 & y_2^2 & y_2^3 & \cdots & y_2^{k-1} & y_2^{k+1} & \cdots & y_2^{n+1} \\ \vdots & \vdots & \vdots & & \vdots & \vdots & & \vdots \\ \vdots & \vdots & \vdots & & \vdots & \vdots & & \vdots \\ 1 & y_n^2 & y_n^3 & \cdots & y_n^{k-1} & y_n^{k+1} & \cdots & y_n^{n+1} \end{vmatrix} \quad (B.2)$$

Also, we define ${}_nW_k(y_1, y_2, \cdots y_n)$ as the elementary symmetric polynomial of degree k among n variables.   We define W=1 when k=0.

For example,
${}_3W_1(y_1, y_2, y_3) \equiv y_1 + y_2 + y_3$
${}_3W_2(y_1, y_2, y_3) \equiv y_1 y_2 + y_2 y_3 + y_3 y_1$

$_nW_{n-1}$ consists of n terms, and is the sum of the products of (n-1) numbers out of n.

$_nW_n$ consists of only one term, which is all products of n variables.

Then, we can prove that :

Formula 1: $G(k, y_1, y_2, \cdots y_n) = {}_nW_{n-k}(y_1, y_2, \cdots y_n) \cdot \prod_{i<j}(y_i - y_j)$ (B.3)

Formula 2: $L(1, k, y_1, y_2, \cdots y_n) = ({}_nW_{n-1} \cdot {}_nW_{n+1-k} - {}_nW_n \cdot {}_nW_{n-k}) \prod_{i<j}(y_i - y_j)$ (B.4)

The G and L are products of Vander monde type determinant and W.

The simplest example of the formula 2 is

$$L(1, 3, y_1, y_2, y_3) = \begin{vmatrix} 1 & y_1^2 & y_1^4 \\ 1 & y_2^2 & y_2^4 \\ 1 & y_3^2 & y_3^4 \end{vmatrix} = (y_1 + y_2)(y_2 + y_3)(y_3 + y_1) \cdot \prod_{i<j}(y_i - y_j)$$

$$= \{(y_1 + y_2 + y_3)(y_1 y_2 + y_2 y_3 + y_3 y_1) - y_1 y_2 y_3\} \cdot \prod_{i<j}(y_i - y_j) = ({}_3W_1 \cdot {}_3W_2 - {}_3W_3 \cdot {}_3W_0) \cdot \prod_{i<j}(y_i - y_j)$$

$_3W_0 = 1$ as defined above.

The most singular pole structure of $H_{N, N-2K}$ is expressed by the ratio

$$\frac{L(1, 2k, y_1, y_2, \ldots y_{N-1})}{G(1, y_1, y_2, \ldots y_{N-1})} \quad .$$

By the formula1 and formula2 above,

$$\frac{L(1, 2k, y_1, y_2, \ldots y_{N-1})}{G(1, y_1, y_2, \ldots y_{N-1})} = \frac{{}_{N-1}W_{N-2} \cdot {}_{N-1}W_{N-2K} - {}_{N-1}W_{N-1} \cdot {}_{N-1}W_{N-1-2K}}{{}_{N-1}W_{N-2}}$$

$$= {}_{N-1}W_{N-2K} - \frac{{}_{N-1}W_{N-1}}{{}_{N-1}W_{N-2}} \cdot {}_{N-1}W_{N-1-2K,} \quad .$$ (B.5)

The ratio $\frac{{}_{N-1}W_{N-1}}{{}_{N-1}W_{N-2}}$ is, when considering the original meaning of the numerator and the denominator going back to the variables $x$, $\frac{1}{\sum_{i=1}^{N-1} x_i}$ which is $-\frac{1}{x_N}$.

Therefore,

$$\frac{L(1, 2k, y_1, y_2, \ldots y_{N-1})}{G(1, y_1, y_2, \ldots y_{N-1})} = {}_{N-1}W_{N-2K} + \frac{1}{x_N} \cdot {}_{N-1}W_{N-1-2K}$$ (B.6)

The meaning of the first term of right hand side, $_{N-1}W_{N-2K}$, is: the sum of the products of $N - 2K$ number of variables chosen out of $N - 1$ variables, excluding $x_M$. There is no $y_N$ or $\dfrac{1}{x_N}$ in the first term.

The meaning of the second term, $\dfrac{1}{x_N} \cdot {}_{N-1}W_{N-1-2K}$, is the such sum of product of $N-2K$ variables in which one of the variables is always $\dfrac{1}{x_N}$.

In total of these two, the meaning of $\dfrac{L(1, 2k, y_1, y_2,...y_{N-1})}{G(1, y_1, y_2,...y_{N-1})}$ is exactly ${}_N W_{N-2K}$ including one more variable $y_N$ which was not contained at the beginning, that is, "Summation of all possible single pole terms $\dfrac{1}{b_1 b_2 .... b_{N-2K}}$, where $b_1, b_2, ....b_{N-2K}$ are chosen $N-2K$ number of variables out of $N$ variables $x_1, x_2, ....x_N$

We considered the case that the second variable of $L$ in the numerator of the ratio $\dfrac{L(1, 2k, y_1, y_2,...y_{N-1})}{G(1, y_1, y_2,...y_{N-1})}$ is even, $2k$, but the same result holds for any of $k$.

Two more observations:

1) Suppose that we start with any set of $N-1$ numbers of variables which satisfies $\sum_{i=1}^{N} x_i = 0$ and calculate the ratio of determinants $\dfrac{L(1, 2k, y_i)}{G(1, y_i)}$. Then, as above, the last variable, say $x_M$, naturally appears from the factor $\dfrac{{}_{N-1}W_{N-1}}{{}_{N-1}W_{N-2}}$. Therefore, in total, the value of $\dfrac{L(1, 2k, y_i)}{G(1, y_i)}$ does not depend on the choice of $N-1$ variables.

Any choice of $N-1$ variables gives the same function $H_{N, N-2K}$.

2) By looking at the second term of right hand side of eq.(B.6) $\dfrac{1}{x_N} \cdot {}_{N-1}W_{N-1-2K}$, we can conclude that, when we regard as if $x_1, x_2,...x_N$ are independent variables,

$$\underset{x_N=0}{\text{Res}}\, H_{N,N-2K}(x_1, x_2,...x_N) = H_{N-1, N-2K-1}(x_1, x_2,...x_{N-1})$$

which is a desirable feature as was explained in ref.[3].

## Appendix C  Higher genus

Let $A_1, A_2, \ldots A_g$ and $B_1, B_2, \ldots B_g$ be a canonical homology basis. We choose canonical holomorphic differentials of the first kind $\omega_1, \omega_2, \ldots \omega_g$ and associated meromorphic differentials of the second kind $r_1, r_2, \ldots r_g$. The periods are given as

$$\oint_{A_I} \omega_J = \delta_{IJ} \tag{C.1}$$

$$\oint_{B_I} \omega_J = \Omega_{IJ} \tag{C.2}$$

$$\oint_{A_I} r_J = \eta_{IJ} \tag{C.3}$$

$$\oint_{B_I} r_J = \eta'_{IJ} \tag{C.4}$$

The followings are examples of genus 2 case. For details, please see [11][12] for example.

Define integrals

$$U_j \equiv \int_{\infty}^{(e_j, 0)} \omega \in C^g \tag{C.5}$$

Then, for g=2, since there are 2g+2 = 6 branch points,

$U_1 = E_1 + \Omega \overline{E_1}$ $\qquad E_1 = \frac{1}{2}(0,0) \quad \overline{E_1} = \frac{1}{2}(1,0)$

$U_2 = E_2 + \Omega \overline{E_2}$ $\qquad E_2 = \frac{1}{2}(1,0) \quad \overline{E_2} = \frac{1}{2}(1,0)$

$U_3 = E_3 + \Omega \overline{E_3}$ $\qquad E_3 = \frac{1}{2}(1,0) \quad \overline{E_3} = \frac{1}{2}(0,1)$

$U_4 = E_4 + \Omega \overline{E_4}$ $\qquad E_4 = \frac{1}{2}(1,1) \quad \overline{E_4} = \frac{1}{2}(0,1)$

$U_5 = E_5 + \Omega \overline{E_5}$ $\qquad E_5 = \frac{1}{2}(1,1) \quad \overline{E_5} = \frac{1}{2}(0,0)$

$U_6 = E_6 + \Omega \overline{E_6}$ $\qquad E_6 = \frac{1}{2}(0,0) \quad \overline{E_6} = \frac{1}{2}(0,0)$

Note that 1+1 = 0.

$\Delta = \Delta^a + \Omega \Delta^b \qquad \Delta^a = E_2 + E_4 = \frac{1}{2}(0,1) \qquad \Delta^b = \overline{E_2} + \overline{E_4} = \frac{1}{2}(1,1)$

Index of 6 odd half periods $\quad \delta_i \equiv U_i + \Delta \quad$ i= 1,2,....6

Index of 10 even half periods $\quad \omega_\delta \equiv U_i + U_j + \Delta \quad 1 \leq i, j \leq 5$

Define the curve as

$$y^2 = 4 \prod_{k=1}^{2g+1} (x - e_k) = R(x) = 4x^{2g+1} + \mu_1 x^{2g} + \mu_2 x^{2g-1} + .... + \mu_{2g+1} \tag{C.6}$$

where a variable y instead of s of eq.(1.3) is used, and one of the branch points $e_{2g+2}$ is fixed at $\infty$.

A polynomial $F(x, z)$ is defined as follows:

$$F(x, z) = \sum_{i=0}^{g} x^i z^i (\mu_{2g-2i}(x+z) + 2\mu_{2g-2i+1})$$

$$= \{(x+z) + 2\mu_1\} x^g z^g + \{\mu_2(x+z) + 2\mu_3\} x^{g-1} z^{g-1} + ... + \mu_{2g}(x+z) + 2\mu_{2g+1} \tag{C.7}$$

The following theorems are known, related to Jacobi's inversion problem

Let $\quad u = \sum_{I=1}^{g} \int_{\infty}^{(x_I, y_I)} \omega \quad$ for any g number of points $(x_1, y_1), ....(x_g, y_g)$, on the curve.

Then the following relations hold, for any r :

$$2y_r = P_{ggg} x_r^{g-1} + P_{gg,g-1} x_r^{g-2} + ..... + P_{gg2} x_r + P_{gg1} \tag{C.8}$$

$$x_r^g - \sum_{J=1}^{g} P_{Jg}(u) x_r^{J-1} = 0 \tag{C.9}$$

$$\sum_{I=1}^{g} \sum_{J=1}^{g} P_{IJ}(u) x_r^{I-1} x_s^{J-1} = \frac{F(x_r, x_s) - 2 y_r y_s}{(x_r - x_s)^2} \tag{C.10}$$

for any choice of two points $(x_r, y_r), (x_s, y_s)$ ,

From these, it can be derived that

$(-1)^{g-J} P_{Jg}(u) = $ *fundamental symmetric function of* $x_1, x_2, ....x_g$ *of order* $g - J + 1$ .

If u is equal to half period $\omega_\delta$, then the point $(x_r, y_r)$ is $(e_r, 0)$.

As an example at g=2, the curve is

$$y^2 = \prod_{k=1}^{5}(x-e_k) = R(x) = x^5 + \mu_1 x^4 + \mu_2 x^3 + .... + \mu_5 \tag{C.11}$$

The function $F(x_1, x_2)$ is given by

$$F(x_1,x_2) = \{(x_1+x_2)+2\mu_1\}x_1^2 x_2^2 + \{\mu_2(x_1+x_2)+2\mu_3\}x_1 x_2 + \mu_4(x_1+x_2)+2\mu_5 \tag{C.12}$$

Suppose that we adopt two points $e_m, e_n$ out of five points $e_1, e_2, .....e_5$.

For a fixed spin structure $\delta$, the equation ( C.10 ) gives only one relationship:

$$P_{11}(\omega_\delta) + P_{12}(\omega_\delta)(e_m + e_n) + P_{22}(\omega_\delta)e_m e_n = \frac{F(e_m, e_n)}{(e_m - e_n)^2} \tag{C.13}$$

On the other hand, eq.(C.9 ) gives

$$e_m^2 - \{P_{12}(\omega_\delta) + P_{22}(\omega_\delta)e_m\} = 0, \quad e_n^2 - \{P_{12}(\omega_\delta) + P_{22}(\omega_\delta)e_n\} = 0 \tag{C.14}$$

Then we have the following solution, at a fixed spin structure,

$$P_{22}(\omega_\delta) = e_m + e_n, \quad P_{12}(\omega_\delta) = P_{21}(\omega_\delta) = -e_m e_n,$$
$$P_{11}(\omega_\delta) = \frac{F(e_m, e_n)}{(e_m - e_n)^2} = (e_p + e_q + e_r)e_m e_n + e_p e_q e_r \tag{C.15}$$

where any of $e_p, e_q, e_r$ is different from $e_m, e_n$

## Appendix D   On $H_{N,N}$ in (2.40) and the function $U(\nu)$ in (2.37)

This appendix is mainly to avoid possible confusing about the definition of $H_{N,N}(x_i)$. In ref.[3], the coefficients $H_{N,M}(x_i)$ are defined in eq.(3.67) and (3.68) in that paper as follows:

$$H_N \equiv [\frac{\nu}{\sigma(\nu)}]^N \prod_{i=1}^{N} \frac{\sigma(x_i + \nu)}{\sigma(x_i)} = \sum_{M=0}^{\infty} H_{N,M}(x_i) \nu^M \tag{D.1}$$

By differentiating M times for both side of this, according to the logic in the text, we have

$$H_{N,M}(x_i) = \frac{1}{M!}\frac{\partial^M}{\partial \nu^M}\exp[\sum_{i=1}^{N}\ln\sigma(x_i+\nu) - \sum_{i=1}^{N}\ln\sigma(x_i) + N\sum_{k=2}^{\infty}\frac{1}{2k}G_{2k}(\tau)\nu^{2k}]\Big|_{\nu=0} \tag{D.2}$$

for all values of M including $M \geq N$. This will match the residue requirement in ref.[3]. $H_{N,N}(x_i)$ is given by just putting M=N in (D.2). By this definition and expression, $H_{4,4}$ has a contribution of $G_4(\tau)$ for example. Apparently $[\frac{v}{\sigma(v)}]^N \prod_{i=1}^{N} \frac{\sigma(x_i+v)}{\sigma(x_i)}$ is given by an infinite sum.

On the other hand, in this document, $[\frac{v}{\sigma(v)}]^N \prod_{i=1}^{N} \frac{\sigma(x_i+v)}{\sigma(x_i)}$ is written as a *finite* sum as in (2.34) or (2.40), expansion by Pe functions:

$$[\frac{v}{\sigma(v)}]^N \prod_{i=1}^{N} \frac{\sigma(x_i+v)}{\sigma(x_i)} = \frac{v^N}{(N-1)!} \sum_{M=0}^{N} H_{N,M}(x_i) P^{(N-2-M)}(v). \tag{2.40}$$

Here, for $M < N$, the coefficients $H_{N,M}(x_i)$ are essentially the same as in (D.2) except the fact that the over-all factor $\frac{1}{M!}$, is replaced with $\frac{(-1)^{N-M}(N-1)!}{(N-1-M)!M!}$ due to simply notational differences, as shown in (2.42):

$$H_{N,M}(x_i) = \frac{(-1)^{N-M}(N-1)!}{(N-1-M)!M!} \frac{\partial^M}{\partial v^M} \exp[\sum_{i=1}^{N} \ln\sigma(x_i+v) - \sum_{i=1}^{N} \ln\sigma(x_i) + N\sum_{k=2}^{\infty} \frac{1}{2k} G_{2k}(\tau) v^{2k}]\big|_{v=0}$$
$$(for \quad M < N) \tag{2.42}$$

We have to pay attention for $H_{N,N}(x_i)$ before differentiating N times. By seeing the expansion of Pe function

$$P(v) = v^{-2} + \sum_{m=1}^{\infty} c_{2m} v^{2m}, \quad c_{2m} = (2m+1) G_{2m+2}(\tau)$$

and its derivative forms, there are many terms in the right hand side of (2.40) which are proportional to $v^N$. All of

$$\frac{v^N}{(N-1)!} \sum_{M=0}^{N} H_{N,M}(x_i)(N-2-M)! \cdot c_{N-2-M}, \tag{D.3}$$

where $N-2-M$ is even, are such terms.

After the differentiation and re-writing, the form of (D.3) will be the last line of the following equation for arbitrary N, M :

$$H_{N,M}(x_i) = \frac{(-1)^{N-M}(N-1)!}{(N-1-M)!M!} \frac{\partial^M}{\partial v^M} \exp[\sum_{i=1}^{N} \ln\sigma(x_i+v) - \sum_{i=1}^{N} \ln\sigma(x_i) + N\sum_{k=2}^{\infty} \frac{1}{2k} G_{2k}(\tau) v^{2k}]\big|_{v=0}$$
$$- \sum_{K=2}^{N-2} \{(K+1)! G_{K+2}(\tau) H_{N,N-K-2}(x_i)\} \delta_{N,M} \tag{D.4}$$

We included this last term $\sum_{K=2}^{N-2} \{(K+1)! G_{K+2}(\tau) H_{N,N-K-2}(x_i)\}\delta_{N,M}$ in the exp function as in (2.44):

$$F(\nu) = \frac{(-1)^{N-M}(N-1)!}{(N-1-M)!M!} \exp[\sum_{i=1}^{N} \ln\sigma(x_i+\nu) - \sum_{i=1}^{N} \ln\sigma(x_i)$$
$$+ N\sum_{k=2}^{\infty} \frac{1}{2k} G_{2k}(\tau)\nu^{2k} - \frac{1}{(N-1)!}\nu^N \sum_{K=2}^{N-2} \{(K+1)! G_{K+2}(\tau) H_{N,N-K-2}(x_i)\}] \quad (2.44)$$

The last term of (2.44) gives non-zero results if and only if M=N and all differentiations with respect to $\nu$ are on this term. Terms of $H_{N,M}(x_i)$ for M<N are obtained by eq. (2.42) before calculating $H_{N,N}(x_i)$ by eq.(2.44).

Note that, the Eisenstein series with highest index, $G_N(\tau)$, does not appear in any of $H_{N,M}(x_i)$, because at M=N the contributions which include $G_N(\tau)$ in the last two terms in (2.44) cancel each other. That is, for example, $H_{4,4}(x_i)$ does not include $G_4(\tau)$, $H_{6,6}(x_i)$ does not include $G_6(\tau)$, etc.

Next we describe the pole structures of the derivatives of the function U defined in (2.37),

$$\frac{\partial^M}{\partial \nu^M} U(\nu) \Big|_{\nu=0} \quad (D.5)$$

where

$$U(\nu) \equiv \exp[\sum_{i=1}^{N} \ln\sigma(x_i+\nu) - \sum_{i=1}^{N} \ln\sigma(x_i)] \quad .. \quad (2.37)$$

The pole structure of (D.5) is the same as that of (2.14). If we select $M$ number of variables $b_1, b_2,....b_M$ out of $N$ number of variables $x_1, x_2,....x_N$, then the poles of (D.5) are

$$\sum_{All\ combinations} \frac{1}{b_1 \cdot b_2 \cdot ....b_M} \quad ,$$

that is, "simultaneous single pole structure" of $M$ number of variables $b_1, b_2,....b_M$" out of $N$ number of variables $x_1, x_2,....x_N$, if appropriate over-all numerical constants which depend on $M$ are multiplied. The proof is done by mathematical induction. For M=1,2,3, it can be seen that this is valid from (2.19), (2.20), (2.21). Suppose this is true for $M$. Then, for one selection of $b_1, b_2, \ldots b_M$, the corresponding term will have the following form of poles before setting $v = 0$:

$$\frac{1}{(b_1+v)(b_2+v)\cdots(b_M+v)} U(v) .$$

By directly differentiating this term with respect to $v$, we have, for the pole part,

$$[-\sum_{L=1}^{M} \frac{1}{(b_L+v)} + \sum_{i=1}^{N} \frac{1}{(x_i+v)}] \frac{1}{(b_1+v)(b_2+v)\cdots(b_M+v)} U(v).$$

That is, from [ ], one more factor $\frac{1}{(x+v)}$ where $x$ is different from any of $b_1, b_2, \ldots b_M$ appears. If we start from another set of $b_j$, the duplicated terms will appear. Those terms will change the numerical factors of each inductively and still the same argument can be applied.

One comment on the poles of spin sum result, eq.(2.47) and (2.48), where we can restrict to the case $M < N$.
The pole structure of (D.5) relates to those of a ratio of two determinants (2.14). The ratio of (2.14) is made from extracting out the most singular part of eq.(2.5). The whole of the structure of (2.5) also has "less singular" poles, by the existence of holomorphic terms in the Pe function and its derivatives in the determinants. The existence of the Eisenstein series $N\sum_{k=2}^{\infty} \frac{1}{2k} G_{2k}(\tau) v^{2k}$ in the exp factor of (2.42) or (2.48) gives such "less singular" parts after the differentiation with respect to $v$. For example, the term $\left.\frac{\partial^{N-2K}}{\partial v^{N-2K}} F_N(x_i, v)\right|_{v=0}$ in eq.(2.47) for K=2,4,5,...contains not only simultaneous single poles of $N-2K$ number of variables but also those of smaller number of variables, as can be seen from (2.48) in general, when $N-2K > 4$.

The pole of $\prod_{i=1}^{N} S_\delta(x_i)$ is the inverse of $N$ product of variables $x_1, x_2, \ldots x_N$.
After summing over spin structures, all poles are still single order for any variable of $x_i$. The number of variables in one pole is equal to or less than $N-4$, as is apparent form eq.(2.47).

# Appendix E  Equivalence of Elliptic Multiple Zeta value method and the method in section 2-6-2  ( Added in the Nov.22 2017 version)

In Appendix E, we clarify the method described in section 2-6-2 on one loop spin sum is equivalent to the results based on elliptic multiple zeta value (eMZV) method described in [4] by some simple observations.    This also gives an explicit proof that the result of spin structure sum in [4] for parity even part is valid for arbitrary N point functions including numerical factors.   It may be useful for non-experts on eMZV methods like me.   In the following it is also described in the spin sum how the notations in eMZV are related to the classical notations in which sigma function and its derivatives are used.

The notations of eMZV necessary here are as follows.
The Eisenstein function and the Kronecker-Eisenstein series are defined by

$$E_j(z,\tau) \equiv \sum_{m,n} \frac{1}{(z+m+n\tau)^j} \quad (E.1)$$

and

$$F(z,\alpha,\tau) \equiv \frac{\theta_1^{(1)}(0)\theta_1(z+\alpha,\tau)}{\theta_1(z,\tau)\theta_1(\alpha,\tau)} \quad (E.2)$$

Eq.(E.1) satisfies the following relations:

$$\frac{\partial \ln \vartheta_1(z,\tau)}{\partial z} = E_1(z,\tau) \qquad \frac{\partial}{\partial z} E_j(z,\tau) = -j E_{j+1}(z,\tau) \quad (E.3)$$

Eq.(E.2) has the expansion form as :

$$F(z,\alpha,\tau) = \frac{1}{\alpha}\exp(-\sum_{j\geq 1}\frac{(-\alpha)^j}{j}(E_j(z,\tau)-G_j(\tau))) \quad (E.4)$$

Some other functions are defined by

$$\Omega(z,\alpha,\tau) \equiv \exp(2\pi i\alpha \frac{\mathrm{Im}(z)}{\mathrm{Im}(\tau)}) F(z,\alpha,\tau) \quad (E.5)$$

$$\alpha\Omega(z,\alpha,\tau) \equiv \sum_{n=0}^{\infty} f^{(n)}(z,\tau)\alpha^n \quad (E.6)$$

$$\Omega_i \equiv \alpha\Omega(x_i,\alpha) \quad (E.7)$$

$$V_p(x_1,x_2,.....,x_N) \equiv (\Omega_1\Omega_2....\Omega_N)\big|_{\alpha^p} \quad (E.8)$$

Note that $V_p(x_1,x_2,.....,x_N)$ is totally different from the quantity $V_{N,N-2K}(x_1,x_2,...x_N)$

defined in (2.8) in the text.

By introducing the differences of inserting points of vertex operators

$x_1 \equiv z_1 - z_2,\ x_2 \equiv z_3 - z_2,\ ....,\ x_N \equiv z_N - z_1$ , which satisfies $\sum_{i=1}^{N} x_i = 0$ ,

we consider the multi-variable function of (E.2), $\prod_{i=1}^{N} F(x_i, \alpha, \tau)$, to consider the spin sum for the N point case.    This equals to

$$\prod_{i=1}^{N} F(x_i, \alpha, \tau) = \prod_{i=1}^{N} \frac{\theta_1^{(1)}(0)\theta_1(x_i + \alpha, \tau)}{\theta_1(x_i, \tau)\theta_1(\alpha, \tau)} = \frac{1}{\alpha^N} \exp(-\sum_{j \geq 1} \frac{(-\alpha)^j}{j}(\sum_{i=1}^{N} E_j(x_i, \tau) - N G_j(\tau)))$$

(E.9)

A key observation is that, by using a standard formula

$$\sigma(z) = \exp(\eta_1 z^2) \frac{\theta_1(z, \tau)}{\theta_1^{(1)}(0)}$$

(E.10)

as well as $\sum_{i=1}^{N} x_i = 0$, $\prod_{i=1}^{N} F(x_i, \alpha, \tau)$ is equal to

$$\prod_{i=1}^{N} F(x_i, \alpha, \tau) = \prod_{i=1}^{N} \frac{\sigma(x_i + \alpha)}{\sigma(x_i)\sigma(\alpha)} \quad .$$

(E.11)

That is, it is equal to Dolan-Goddard generating function (2.25), replaced the auxiliary variable $\nu$ with $\alpha$.

Therefore, all arguments in section 2-6-2 can be applied afterwards.    Or more directly, we modify the single $F(z, \alpha, \tau)$ in eq.(E.4) ,    by using

$$E_j(z, \tau) = (-1)^{j-1} \frac{1}{(j-1)!} \frac{\partial^j \ln \vartheta_1(z, \tau)}{\partial z^j} \quad ,$$

(E.12)

as:

$$F(z, \alpha, \tau) = \frac{1}{\alpha} \exp(-\sum_{j \geq 1} \frac{(-\alpha)^j}{j}(E_j(z, \tau) - G_j(\tau))) = \frac{1}{\alpha} \exp(\sum_{j \geq 1} \frac{\alpha^j}{j!} \frac{\partial^j}{\partial z^j} \ln \theta_1(z, \tau) + \sum_{k=1}^{\infty} \frac{\alpha^{2k}}{2k} G_{2k}(\tau))$$

$$= \frac{1}{\alpha} \exp(\ln \theta_1(z + \alpha, \tau) - \ln \theta_1(z, \tau) + \sum_{k=1}^{\infty} \frac{\alpha^{2k}}{2k} G_{2k}(\tau))$$

$$= \frac{1}{\alpha} \exp(\ln \sigma(z + \alpha, \tau) - \ln \sigma(z, \tau) + \sum_{k=2}^{\infty} \frac{\alpha^{2k}}{2k} G_{2k}(\tau))$$

(E.13)

In the last equality, we used the k=1 term in the exp function $\frac{\alpha^2}{2} G_2(\tau)$ to express $\ln \theta_1(z, \tau)$ by the sigma function.    Therefore, we have

$$\alpha^N \prod_{i=1}^{N} F(x_i,\alpha,\tau) = \exp(\sum_{i=1}^{N} \ln\sigma(x_i+\alpha) - \sum_{i=1}^{N}\ln\sigma(x_i) + N\sum_{k=2}^{\infty}\frac{\alpha^{2k}}{2k}G_{2k}(\tau)) \qquad (E.14)$$

which is the same as the (2.39), by replacing $\alpha$ with $v$ :

$$[\frac{v}{\sigma(v)}]^N \prod_{i=1}^{N}\frac{\sigma(x_i+v)}{\sigma(x_i)} = \exp[\sum_{i=1}^{N}\ln\sigma(x_i+v) - \sum_{i=1}^{N}\ln\sigma(x_i) - N\sum_{m,n}[\ln(1-\frac{v}{\Omega_{m,n}}) + (\frac{v}{\Omega_{m,n}} + \frac{v^2}{2\Omega_{m,n}^2})]]$$

$$(2.39)$$

$$= \exp[\sum_{i=1}^{N}\ln\sigma(x_i+v) - \sum_{i=1}^{N}\ln\sigma(x_i) + N\sum_{k=2}^{\infty}\frac{v^{2k}}{2k}G_{2k}(\tau)]$$

On the other hand, from (E.5) (E.7) (E.8) as well as $\sum_{i=1}^{N}x_i = 0$ , we have

$$V_p(x_1,x_2,\ldots,x_N) = \frac{1}{p!}\frac{\partial^P}{\partial\alpha^P}[\alpha^N\prod_{i=1}^{N}F(x_i,\alpha,\tau)]\Big|_{\alpha=0} \qquad (E.15)$$

which is the coefficient of Taylor expansion of a holomorphic function $\alpha^N\prod_{i=1}^{N}F(x_i,\alpha,\tau)$ by (E.8).

For convenience, without fear of repetition, we describe in the following how these observations are related to the argument of summing over spin structures given in 2-6-2.

We regard the product $\prod_{i=1}^{N}F(x_i,\alpha,\tau)$ as an elliptic function of $\alpha$, and expand by the derivatives of Pe function as

$$\prod_{i=1}^{N}F(x_i,\alpha,\tau) = \sum_{M=0}^{N}h_{N,M}(x_i)P^{(N-2-M)}(\alpha) \qquad (E.16)$$

where $h_{N,M}(x_i)$ are expansion coefficients defined here. Considering the poles of Pe :

$$P(\alpha) = \alpha^{-2} + \sum_{m=1}^{\infty}(2m+1)G_{2m+2}(\tau)\alpha^{2m}$$

$$P^{(n)}(\alpha) = \frac{d_n}{\alpha^{n+2}} + n!(2m+1)G_{2m+2}(\tau) + O(\alpha)\ldots\ldots$$

$$d_n = (-1)^n(n+1)! \quad , \qquad (E.17)$$

and multiplying $\alpha^N$ on both sides of (E.16), differentiating M times with respect to

$\alpha$, we have the explicit form of $h_{N,M}(x_i)$:

$$h_{N,M}(x_i) = \frac{1}{M! d_{N-2-M}} \frac{\partial^M}{\partial \alpha^M} [\alpha^N \prod_{i=1}^{N} F(x_i, \alpha, \tau)] \big|_{\alpha=0} \qquad (E.18)$$

which is valid for $M < N$.

Once the coefficients $h_{N,M}(x_i)$ are obtained, we can have the N product of fermion correlation functions $\prod_{i=1}^{N} S_\delta(x_i)$ under the condition $\sum_{i=1}^{N} x_i = 0$ by setting $\alpha$ equals to half period $\omega_\delta$ by (2.27), (2.32) and (E.11):

$$\prod_{i=1}^{N} S_\delta(x_i) = \prod_{i=1}^{N} F(x_i, \omega_\delta, \tau) = \sum_{M=0}^{N} h_{N,M}(x_i) P^{(N-2-M)}(\omega_\delta)$$

$$= \sum_{M=0}^{N} \frac{(-1)^{N-2-M}}{M!(N-M-1)!} \frac{\partial^M}{\partial \alpha^M} [\alpha^N \prod_{i=1}^{N} F(x_i, \alpha, \tau)] \big|_{\alpha=0} P^{(N-2-M)}(\omega_\delta) \qquad (E.19)$$

By comparing (E.15) and (E.18), we have

$$\prod_{i=1}^{N} S_\delta(x_i) = \sum_{M=0}^{N} \frac{(-1)^{N-2-M}}{(N-M-1)!} V_M(x_1, x_2, \ldots, x_N) P^{(N-2-M)}(\omega_\delta) \qquad (E.20)$$

and further, since $P^{(ODD)}(\omega_\delta) = 0$ and $Q_K(e_\delta) = P^{(2K-2)}(\omega_\delta)$ as in (1.12),

$$\prod_{i=1}^{N} S_\delta(x_i) = \sum_{K=0}^{\left[\frac{N}{2}\right]} \frac{1}{(2K-1)!} V_{N-2K}(x_1, x_2, \ldots, x_N) Q_K(e_\delta) \quad . \qquad (E.21)$$

And

$$V_p(x_1, x_2, \ldots, x_N) = \frac{1}{p!} \frac{\partial^p}{\partial \alpha^p} \exp(\sum_{i=1}^{N} \ln \sigma(x_i + \alpha) - \sum_{i=1}^{N} \ln \sigma(x_i) + N \sum_{k=2}^{\infty} \frac{\alpha^{2k}}{2k} G_{2k}(\tau))\big|_{\alpha=0}$$

(for $p < N$). $\qquad (E.22)$

To have spin sum of even parity, we rewrite the theta constants $(-1)^\delta \left[\frac{\theta_{\delta+1}(0)}{\theta_1^{(1)}(0)}\right]^4$ by branch points as $\frac{(e_1 - e_3)}{\sqrt{D}}, \frac{(e_3 - e_2)}{\sqrt{D}}, \frac{(e_2 - e_1)}{\sqrt{D}}$ where $\sqrt{D} \equiv \prod_{i<j}^{g}(e_i - e_j)$, and as in (2.47), we calculate to have

$$\sum_{\delta=1,2,3} (-1)^\delta \left[ \frac{\theta_{\delta+1}(0)}{\theta_1^{(1)}(0)} \right]^4 \prod_{i=1}^N S_\delta(x_i) = \sum_{K=0}^{\left[\frac{N}{2}\right]} E_K(\tau) V_{N-2K}(x_1, x_2, ... x_N) \tag{E.23}$$

with $E_0(\tau) = E_1(\tau) = E_3(\tau) = 0$, $E_2(\tau) = 1$ by the algebra of $e_\delta$.

Here $E_K(\tau)$ is not the Eisenstein function but the polynomial defined in (2.48):

$$E_K(\tau) \equiv \frac{1}{(2K-1)!} \frac{(e_1 - e_3)Q_K(e_2) + (e_3 - e_2)Q_K(e_1) + (e_2 - e_1)Q_K(e_3)}{(e_1 - e_3)(e_3 - e_2)(e_2 - e_1)} \tag{2.48}$$

and $(2K-1)!$ is equal to the numerical coefficient of $P^K$ in the polynomial $Q_K$ which is the highest degree term of the polynomial. The whole of $E_K(\tau)$ can always be represented by the elementary symmetric functions of $e_\delta$ as is easily proved, and therefore by the Eisenstein series. Note again that $E_0(\tau) = 0$ ensures that we do not need the form of $V_p(x_1, x_2, ....., x_N)$ for $p = N$.

Some notes:
1)
In eMZV, the basic building blocks of the theory are functions $f^{(n)}(z, \tau)$, defined in the expansion form of (E.6). The functions $V_p(x_1, x_2, ....., x_N)$ defined in (E.8) are expressed by $f^{(n)}(z, \tau)$, and give the results of spin sum, by (E.23). This is based on a wide framework described in [4], and is convenient for general theoretical considerations of the results. $V_p(x_1, x_2, ....., x_N)$ are also related to the classical notations in which sigma function and its derivatives are used, as in (E.22). In (E.22), the Eisenstein series appear before considering spin structures, because $f^{(n)}(z, \tau)$ naturally has such modular forms inside. For example, Dolan Goddard pointed out the additional $k_4$ term, eq(3.52) in [3]. This is the $G_4(\tau)$ in $N \sum_{k=2}^\infty \frac{\alpha^{2k}}{2k} G_{2k}(\tau)$ of (E.22), and it is irrespective of spin structures, that is, it is not from $E_K(\tau)$ factors in (E.23).

We have to do total p times of differentiations in (E.22) with respect to the auxiliary variable $\alpha$. We may do 0,4,6,8,… times of differentiations to give the terms proportional to the Eisenstein series on the terms $N\sum_{k=2}^{\infty}\frac{\alpha^{2k}}{2k}G_{2k}(\tau)$. The rest $p-2L$, $L=0,2,3,4,...$ times of differentiations are on $\exp(\sum_{i=1}^{N}\ln\sigma(x_i+\alpha)-\sum_{i=1}^{N}\ln\sigma(x_i))$. This gives "simultaneous single pole structure" of $p-2L$ number of variables chosen out of $N$ number of variables $x_1, x_2, ....x_N$, after setting $\alpha=0$, as described in Appendix D. This makes the pole structures of the spin sum results clear.

The total of $V_p(x_1, x_2, ....., x_N)$ is the summation of all such poles of different orders of $p$, $p-4$, $p-6$, .... .

2)

The fundamental reason why $V_p(x_1, x_2, ....., x_N)$ is related to the spin structures comes from the following relations under the condition $\sum_{i=1}^{N} x_i = 0$ :

$$\prod_{i=1}^{N} S_\delta(x_i) = \prod_{i=1}^{N} F(x_i, \omega_\delta, \tau) = \sum_{M=0}^{N} h_{N,M}(x_i) P^{(N-2-M)}(\omega_\delta) \tag{E.19}$$

as well as the comparison of $V_p(x_1, x_2, ....., x_N)$ , (E.15) with $h_{N,M}(x_i)$ , (E.18).

As one more example, we show a spin sum calculation of quarter-maximal supersymmetric case. Note that, this result is already obtained in [5] in a general manner in eMZV language, and the calculation below is only re-writing their result in classical notations.

We calculate

$$\sum_{\delta=1,2,3}(-1)^\delta\left[\frac{\theta_{\delta+1}(0)}{\theta_1^{(1)}(0)}\right]^4 \prod_{j=1}^{3} S_\delta(\gamma_j) \prod_{i=1}^{N} S_\delta(x_i) \tag{E.24}$$

for $\sum_{i=1}^{N} x_i = 0$ and $\sum_{j=1}^{3} \gamma_j = 0$ in the notations of [5].

Here instead of calculating N+3 factors of Szego kernels, we evaluate $\prod_{j=1}^{3} S_\delta(\gamma_j)$ first, by (E.21) and (E.22),

$$\prod_{j=1}^{3} S_\delta(\gamma_j) = V_3(\gamma_1, \gamma_2, \gamma_3) + V_1(\gamma_1, \gamma_2, \gamma_3)e_\delta$$

$$= \frac{1}{6}\{\left[\sum_{i=1}^{3}\zeta(\gamma_i)\right]^3 - 3\left[\sum_{i=1}^{3}\zeta(\gamma_i)\right]\left[\sum_{i=1}^{3}P(\gamma_i)\right] - \left[\sum_{i=1}^{3}P^{(1)}(\gamma_i)\right]\} + \left[\sum_{i=1}^{3}\zeta(\gamma_i)\right]e_\delta \quad .$$
(E.25)

Note that, (E.22) is valid only for $p < N$, but $V_3(\gamma_1, \gamma_2, \gamma_3)$ is the case $p = N = 3$. Going back to (D.3) and (D.4), we can say that for $N = 3$, the additional term in the last line of (D.4) is not necessary.

By using $\left[\sum_{i=1}^{3}\zeta(\gamma_i)\right]^2 = \sum_{i=1}^{3}P(\gamma_i)$ for the three variables case with $\sum_{j=1}^{3}\gamma_j = 0$, we have

$$\prod_{i=1}^{3}S_\delta(\gamma_i) = -\frac{1}{3}\left[\sum_{i=1}^{3}\zeta(\gamma_i)\right]^3 - \frac{1}{6}\left[\sum_{i=1}^{3}P^{(1)}(\gamma_i)\right] + \left[\sum_{i=1}^{3}\zeta(\gamma_i)\right]e_\delta \quad \text{(E.26)}$$

Therefore,

$$\sum_{\delta=1,2,3}(-1)^\delta \left[\frac{\theta_{\delta+1}(0)}{\theta_1^{(1)}(0)}\right]^4 \prod_{j=1}^{3}S_\delta(\gamma_j) \prod_{i=1}^{N}S_\delta(x_i) = A_3 + B_3 \quad \text{(E.27)}$$

where

$$A_3 = \{-\frac{1}{3}\left[\sum_{i=1}^{3}\zeta(\gamma_i)\right]^3 - \frac{1}{6}\left[\sum_{i=1}^{3}P^{(1)}(\gamma_i)\right]\}\sum_{K=0}^{\left[\frac{N}{2}\right]}E_K(\tau)V_{N-2K}(x_1, x_2,...x_N) \quad \text{(E.28)}$$

$$B_3 = \left[\sum_{i=1}^{3}\zeta(\gamma_i)\right]\sum_{K=0}^{\left[\frac{N}{2}\right]}\overline{E_K}(\tau)V_{N-2K}(x_1, x_2,...x_N) \quad \text{(E.29)}$$

The result of $A_3$ is proportional to the maximal supersymmetric case (E.23) with a factor $E_K(\tau)$ defined in (2.48), whereas $B_3$ contains a modified factor:

$$\overline{E_K}(\tau) \equiv \frac{1}{(2K-1)!}\frac{(e_1-e_3)e_2Q_K(e_2) + (e_3-e_2)e_1Q_K(e_1) + (e_2-e_1)e_3Q_K(e_3)}{(e_1-e_3)(e_3-e_2)(e_2-e_1)} \quad . \quad \text{(E.30)}$$

This can also be represented by the Eisenstein series in general. As can be calculated easily, $\overline{E_0}(\tau) = \overline{E_2}(\tau) = 0$, $\overline{E_1}(\tau) = 1$ by the algebra of $e_\delta$.

$V_{N-2K}(x_1, x_2,...x_N)$ are derivatives of the generating function as in (E.22), nothing

changed from the maximal supersymmetric case.

The same way can be applied to the half maximal case.    We evaluate

$$\sum_{\delta=1,2,3} (-1)^\delta \left[ \frac{\theta_{\delta+1}(0)}{\theta_1^{(1)}(0)} \right]^4 S_\delta(kv)\, S_\delta(-kv) \prod_{i=1}^N S_\delta(x_i) \quad \text{for } \sum_{i=1}^N x_i = 0.$$

Since $\zeta(kv)+\zeta(-kv)=0$,   we have

$$\sum_{\delta=1,2,3} (-1)^\delta \left[ \frac{\theta_{\delta+1}(0)}{\theta_1^{(1)}(0)} \right]^4 S_\delta(kv)\, S_\delta(-kv) \prod_{i=1}^N S_\delta(x_i) = A_2 + B_2 \qquad (E.31)$$

where

$$A_2 = -P(kv) \sum_{K=0}^{\left[\frac{N}{2}\right]} E_K(\tau)\, V_{N-2K}(x_1, x_2, \ldots x_N) \qquad (E.32)$$

$$B_2 = \sum_{K=0}^{\left[\frac{N}{2}\right]} \overline{E_K(\tau)}\, V_{N-2K}(x_1, x_2, \ldots x_N) \qquad (E.33)$$

A.G.  Tsuchiya   E-mail Address   colasumi@theia.ocn.ne.jp